\documentclass[%
 aip,
 amsmath,amssymb,
 reprint,%
]{revtex4-1}

\usepackage{graphicx}
\usepackage{dcolumn}
\usepackage{bm}

\usepackage[utf8]{inputenc}
\usepackage[T1]{fontenc}
\usepackage{mathptmx}
\usepackage{etoolbox}
\usepackage{xcolor}
\usepackage{multirow}
\usepackage{subcaption}
\newcommand*{\SHs}{HgSH$^{+}$}
\newcommand*{\SHd}{Hg(SH)$_{2}$}
\newcommand*{\SHt}{[Hg(SH)$_{3}$]$^-$}
\newcommand*{\SHq}{[Hg(SH)$_4$]$^{2-}$}
\newcommand*{\SMed}{Hg(SMe)$_2$}
\newcommand*{\SPhd}{Hg(SPh)$_2$}
\newcommand*{\Cysd}{Hg(Cys)$_2$}

\newcommand*{\SEtd}{Hg(SEt)$_2$}
\newcommand*{\asymcys}{Hg(SH)(Cys)}
\newcommand*{\Med}{Hg(CH$_3$)$_2$}
\usepackage{braket}

\makeatletter
\def\@email#1#2{%
 \endgroup
 \patchcmd{\titleblock@produce}
  {\frontmatter@RRAPformat}
  {\frontmatter@RRAPformat{\produce@RRAP{*#1\href{mailto:#2}{#2}}}\frontmatter@RRAPformat}
  {}{}
}%
\makeatother
\begin{document}

\preprint{AIP/123-QED}

\title[Geometry dependence of the chemical shift of mercury]{On the Geometry Dependence of the NMR Chemical Shift of Mercury in Thiolate Complexes: A Relativistic DFT Study}
\author{Haide Wu}

\altaffiliation[Present Address: ]{Department of Chemistry, Aarhus University, Denmark}
\affiliation{Department of Chemistry, University of Copenhagen, Universitetsparken 5, DK-2100 Copenhagen, Denmark}
\author{Lars Hemmingsen}%
\affiliation{Department of Chemistry, University of Copenhagen, Universitetsparken 5, DK-2100 Copenhagen, Denmark}
\author{Stephan P. A. Sauer}%
\email{sauer@chem.ku.dk}
\affiliation{Department of Chemistry, University of Copenhagen, Universitetsparken 5, DK-2100 Copenhagen, Denmark}


\date{\today}

\begin{abstract}
Thiolate containing mercury(II) complexes of the general formula [Hg(SR)$_n$]$^{2-n}$ have been of great interest since the toxicity of mercury was recognized. $^{199}$Hg nuclear magnetic resonance spectroscopy (NMR) is a powerful tool for characterization of mercury complexes. In this work, the Hg shielding constants in a series of [Hg(SR)$_n$]$^{2-n}$ complexes are therefore investigated computationally with particular emphasis on their geometry dependence. Geometry optimizations and NMR chemical shift calculations are performed at the density functional theory (DFT) level with both the zeroth-order regular approximation (ZORA) and four-component relativistic methods. The four exchange-correlation (XC) functionals PBE0, PBE, B3LYP and BLYP are used in combination with either Dyall's Gaussian-type (GTO) or Slater-type orbitals (STOs) basis sets. Comparing ZORA and four-component calculations, one observes that the calculated shielding constants for a given molecular geometry have a constant difference of $\sim$1070 ppm. This confirms that ZORA is an acceptable relativistic method to compute NMR chemical shifts. The combinations of 4-component/PBE0/v3z and ZORA/PBE0/QZ4P are applied to explore the geometry dependence of the isotropic shielding. For a given coordination number the distance between mercury and sulfur is the key factor affecting the shielding constant, while changes in bond and dihedral angles and even different side groups have relatively little impact.   
\end{abstract}

\maketitle

\section{Introduction}
Molecules with sulfhydryl groups are referred to as thiols or mercaptans.\cite{ajsuvakova2020sulfhydryl} 
The affinity of mercury(II) for thiolates is even higher than that of other thiophilic metal ions such as Cd(II) and Pb(II).\cite{mutter2007comments} 
Since sulfhydryl (SH) groups are ubiquitous and most of them are important for the function or structure of numerous proteins,\cite{nordberg2014handbook} thiol reactivity plays an important role in mercury toxicity. 

Nuclear magnetic resonance (NMR) provides in general a means for characterization of molecular structure and dynamics. 
The sensitivity of $^{199}$Hg ($I=\frac{1}{2}$, natural abundance=16.8\%) chemical shifts to the primary coordination sphere of mercury complexes makes $^{199}$Hg NMR a powerful tool for elucidating metal-binding sites in proteins.\cite{utschig1995mercury} 

For calculation on systems containing heavy elements, such as mercury, a proper treatment of the relativistic effects has to be introduced. 
From a computational point of view, there are several ways of treating relativistic effects. 
The first option is fully relativistic four-component (4-comp) linear response calculations.\cite{visscher1999full,iliavs2009gauge,gomez2002fully,vaara2003relativistic,lantto2006relativistic}
An alternative option are the computationally less demanding two-component methods\cite{chang1986regular,lenthe1993relativistic,van1994relativistic,van1996zero,van1996relativistic,wolff1999density} such as the zeroth-order regular approximation (ZORA) method. 
The usefulness of ZORA calculations of NMR parameters for heavy elements has been shown and reviewed in many studies.\cite{Autschbach2004c, Autschbach2004d, Kaupp2004, Buhl2004b, Autschbach2009a, Autschbach20120489, GasPhaseNMR-Ch-8-267-2016-Repisky, Vicha2020, arcisauskaite2011nuclear, spas155, spas193, spas204, spas212} 
\citeauthor{arcisauskaite2011nuclear} e.g. have previously investigated relativistic effects on Hg chemical shifts in mercury halide compounds. \cite{arcisauskaite2011nuclear} A comparison between three methods were reported: the fully relativistic four-component approach, linear response elimination of small component (LR-ESC) \cite{Melo2003,nmr04-jcp121-6798} and ZORA. 
They confirmed that chemical shifts calculated by ZORA may be adequate.

Mercury can form two- three- and four-coordinated complexes with cysteine,\cite{utschig1995mercury,jalilehvand2006mercury} sulfur bridges are possible \cite{shang2011mercury} and Hg may exchange between coordinating groups.\cite{ajsuvakova2020sulfhydryl} 
Large side groups on the sulfur atoms may cause geometrical distortion of the complex, resulting different coordinate bond length or bond angle than complexes with small ligands.\cite{steele1997structures, santos1991solid}
The dependence of the $^{199}$Hg chemical shifts  on geometry can be a very useful tool to investigate these effects. 

In this study we systematically investigate the effect of changing the geometry of mercury-thiolate-complexes, [Hg(SR)$_n$]$^{2-n}$. 
Before doing so we will (1) compare the results of ZORA and fully relativistic four-component calculations not only for the shielding constants and chemical shifts but also for the geometry of the [Hg(SR)$_n$]$^{2-n}$ complexes; (2) investigate the dependence of the calculated shielding constants on basis set and XC-functionals.

\section{Computational Details}
\subsection{Choice of Model Systems}
For the investigation of the effect of basis sets, of the XC-functionals and of the relativistic method on the geometry optimizations and shielding calculations, we have studied the simple \SHs, \SHd, \SHt and \SHq systems. 
For the following study of the geometry dependence of the Hg shielding constant we included also the Hg(SR)$_2$ complexes with the side groups "R" being methyl (Me), ethyl (Et), phenyl (Ph), cysteine (Cys).

\subsection{Four-component Calculations}
The four-component relativistic calculations with the Dirac-Coulomb Hamiltonian were carried out using the DIRAC program\cite{DIRAC21} at density functional theory (DFT) level with the general gradient approximation (GGA) exchange-correlation functionals BLYP \cite{miehlich1989results,becke1988density} and PBE \cite{perdew1996generalized,perdew1998perdew} and the hybrid GGA XC-functionals B3LYP \cite{becke1993becke, RN1} and PBE0. \cite{adamo1999toward,perdew1996rationale} 
In these calculations Dyall's v3z basis set \cite{dyall2011relativistic} was applied on all atoms. 
For the basis set study with the PBE0 functional, also Dyall's v2z was employed as well as the "core-valence" and "all-electron" versions of the v3z basis set, i.e. cv3z and ae3z. 
The core-valence basis sets include the (n-2) shell for the s-elements, the (n-1) shell for the p-elements, the (n-1) shell for the d-elements.
The all-electron basis sets include correlating functions for all shells, down to the 1s for all elements. 
Symmetry was not enforced in the calculations through out this article. 

\subsection{Spin-Orbitals ZORA Calculations}
The spin-orbit ZORA calculations were performed with the Amsterdam Density Functional module (ADF) \cite{ADF2001, ADF2017authors, guerra1998towards} of the Amsterdam Modeling Suite (AMS). 
ZORA adapted DZP, TZP, TZ2P and QZ4P Slater-type orbital basis sets were applied \cite{chong2004even, chong2005augmenting, van2003optimized} in combination with the PBE0 and B3LYP XC-functionals for both geometry optimizations and calculations of the nuclear magnetic shielding constants.\cite{krykunov2009hybrid} 
A spherical Gaussian nuclear charge distribution model \cite{autschbach2009magnitude} was applied. The FXC option \cite{autschbach2013role} was activated for the shielding calculations in order to account correctly for the response of the DFT exchange-correlation potential to the external magnetic field perturbation, $f_{XC}$. 

\subsection{XC-Functional and Basis Set Dependence Investigation}
In the study on the dependence of our results on basis sets and XC-functional, we firstly optimized molecular geometries (including reference compound \Med) with specific basis sets and XC-functional. Afterwards, NMR shielding calculations were proceeded based on the resulted geometry consistently with same XC-functional, basis sets and relativistic treatment. Results by this workflow will show which method would provide us more convincible results.

\subsection{Geometry Dependence Investigation}
In the investigation of the geometry effect, both geometry optimizations and NMR shielding calculations were performed at SO-ZORA/DFT level with PBE0 functional. QZ4P basis set was applied on mercury and sulfur atoms while TZ2P on carbon atoms and DZP on all other atoms. Starting from the the optimized geometry, modifications were introduced by setting specific geometry parameters, $e.g.$ bond length(s), bond angle(s) or dihedral(s), to a certain value. With the modified geometry parameter(s) fixed, constrained geometry optimizations were carried out. NMR shielding calculations were perform at the same level of theory by using the constrained optimized geometries. For a further comparison between relativistic methods, four-component calculations of small molecules, \SHd, \SHt, Hg(SMe)$_{2}$ and Hg(SEt)$_{2}$, were also perform based at the constrained optimized geometries.

\section{Results and Discussion}
\begin{figure}
\centering
\begin{subfigure}[b]{0.2\textwidth}
  \centering
  \includegraphics[width=1\textwidth]{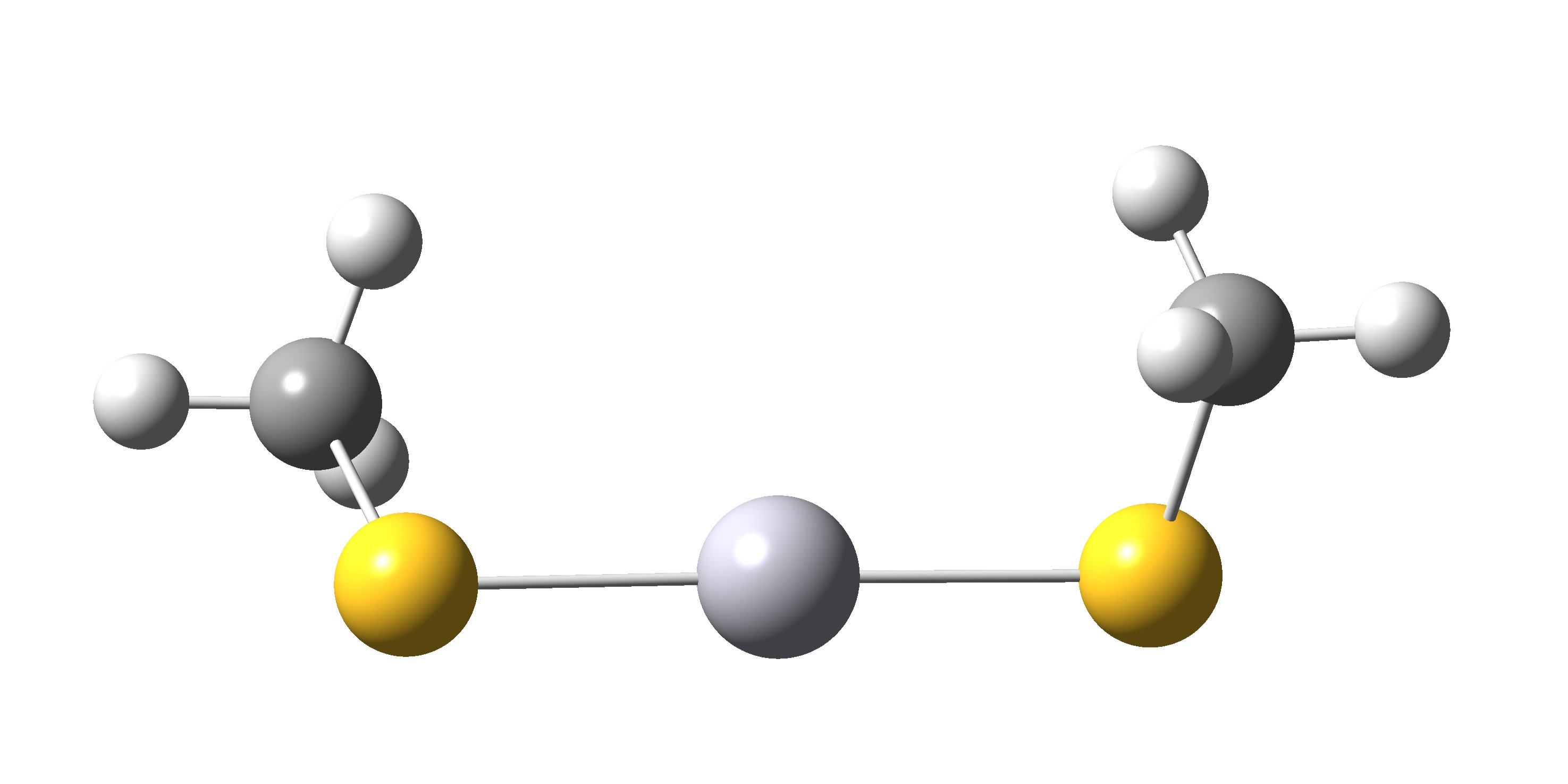}
  \caption{\label{HgMe2}\Med} 
\end{subfigure}
\hfill
\begin{subfigure}[b]{0.2\textwidth}
  \centering
  \includegraphics[width=1\textwidth]{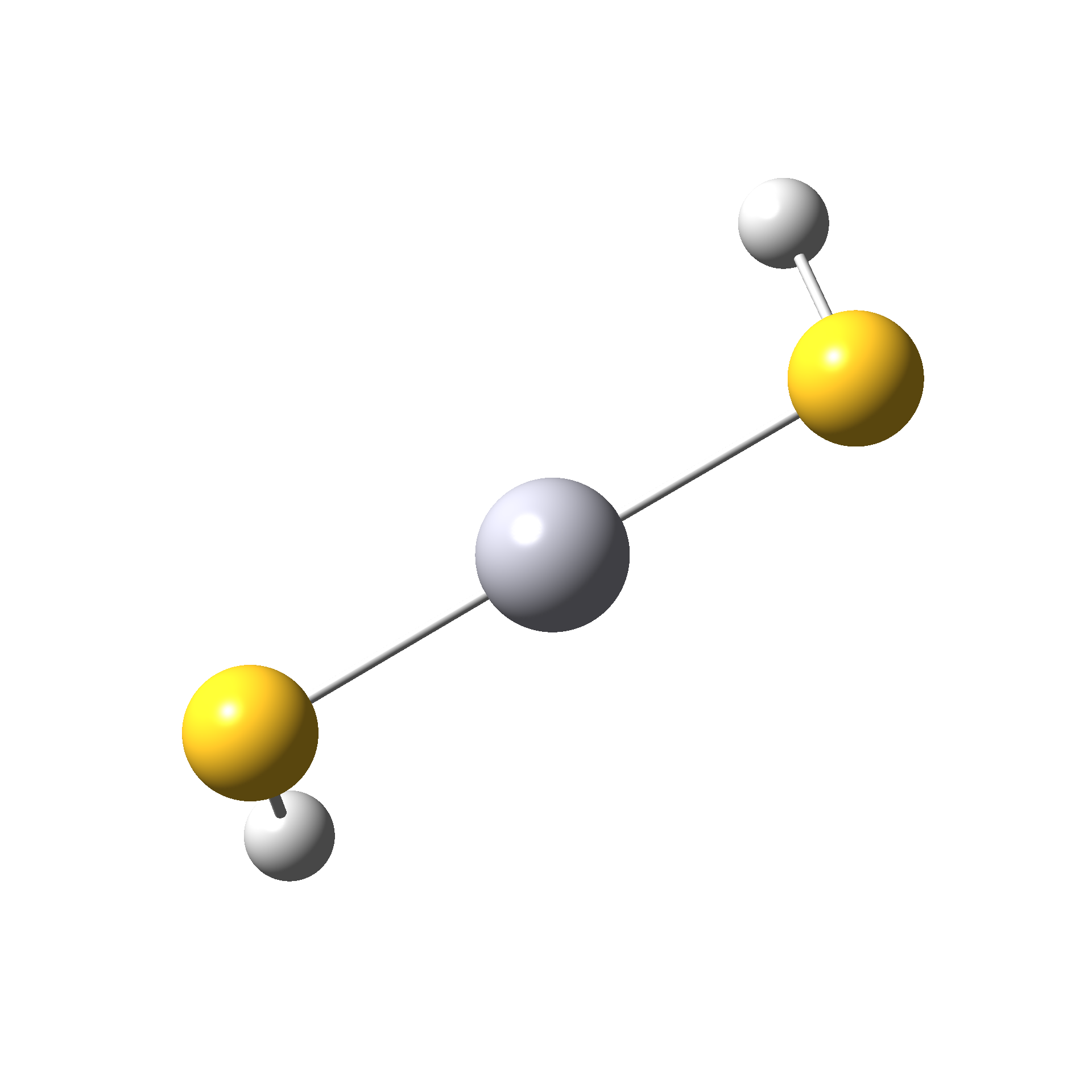}
  \caption{\label{HgSH2}\SHd} 
\end{subfigure}
\hfill
\begin{subfigure}[b]{0.2\textwidth}
  \centering
  \includegraphics[width=1\textwidth]{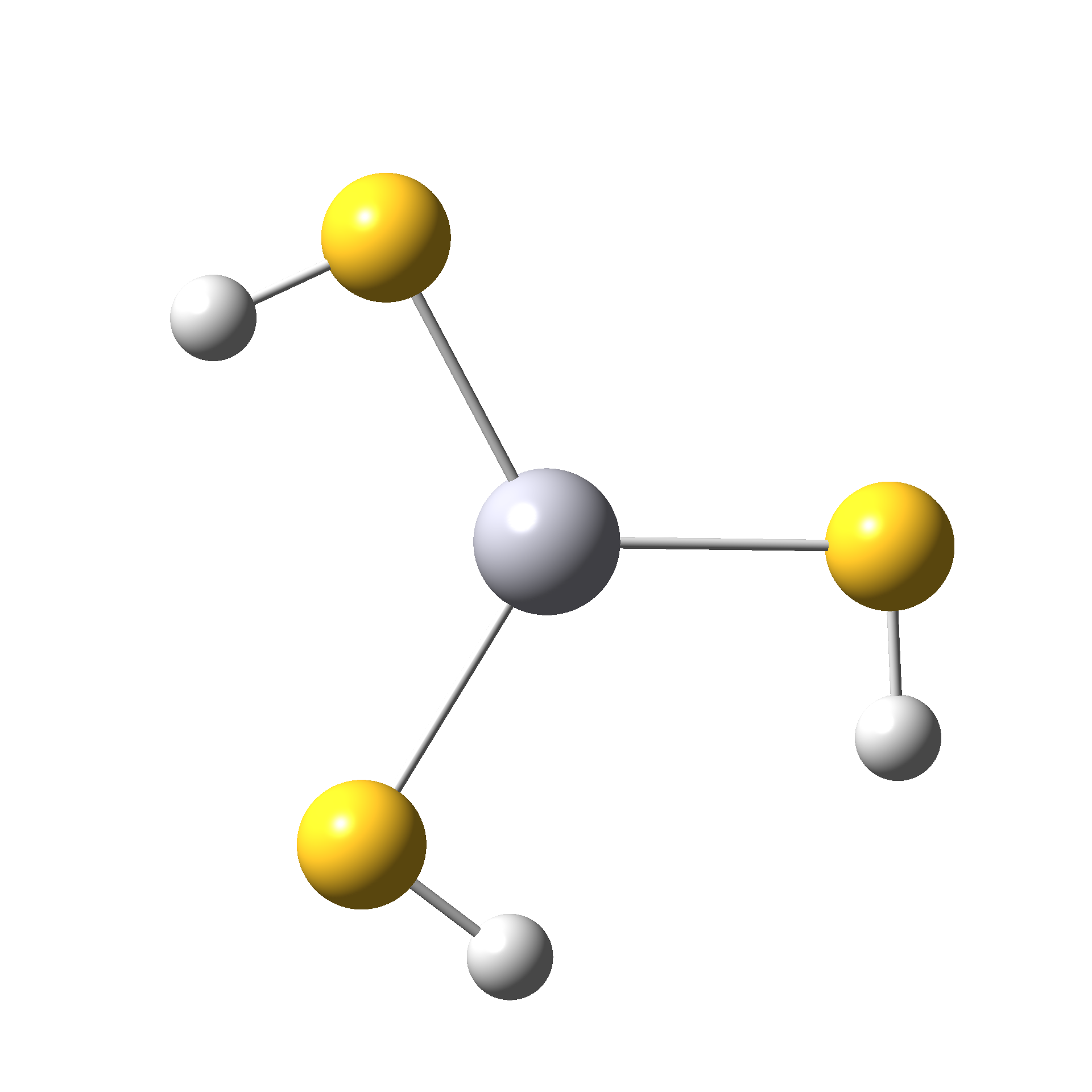}
  \caption{\label{HgSH3}\SHt} 
\end{subfigure}
\hfill
\begin{subfigure}[b]{0.2\textwidth}
  \centering
  \includegraphics[width=1\textwidth]{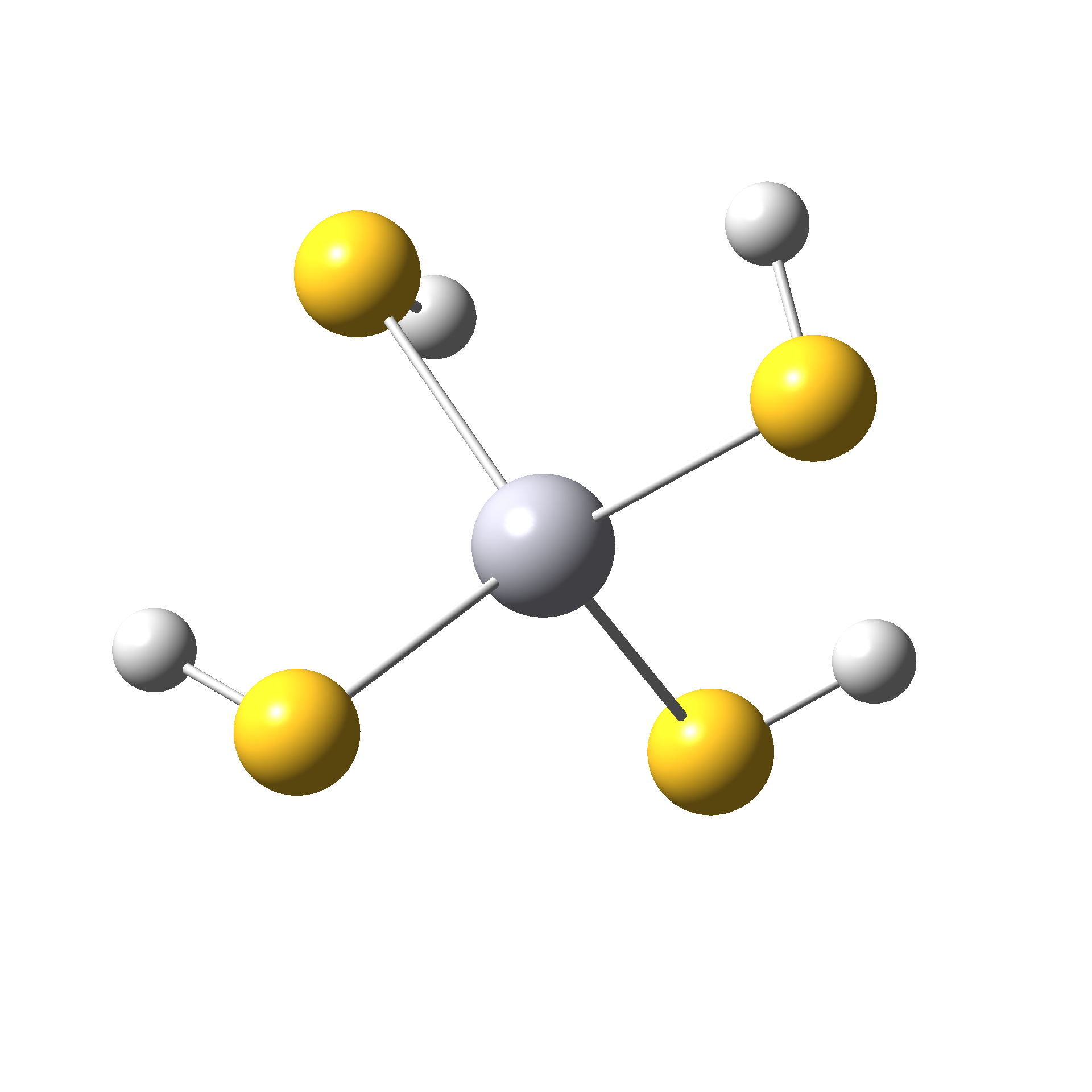}
  \caption{\label{HgSH4}\SHq} 
\end{subfigure}
\label{fig:molecular geometries}
\caption{Geometries of \Med and [Hg(SH)$_n$]$^{2-n}$, optimized at SO-ZORA/QZ4P level of theory.}
\end{figure}
\subsection{Geometry Optimization}
First we investigated the dependence of the optimized geometries on the employed basis sets and XC-functionals. Since we were mainly interested in the chemical shifts of mercury, the chemical environment around mercury was our focus. Calculated bond lengths and bond angles between mercury, sulfur and hydrogen ($R\rm_{S-Hg}$, $R\rm_{S-H}$, $\gamma\rm_{S-Hg-S}$ and $\theta\rm_{H-S-Hg}$) will therefore be compared in this section.

\subsubsection{Dependence on Basis Sets}
\begin{figure}
    \centering
    \includegraphics[width=0.45\textwidth]{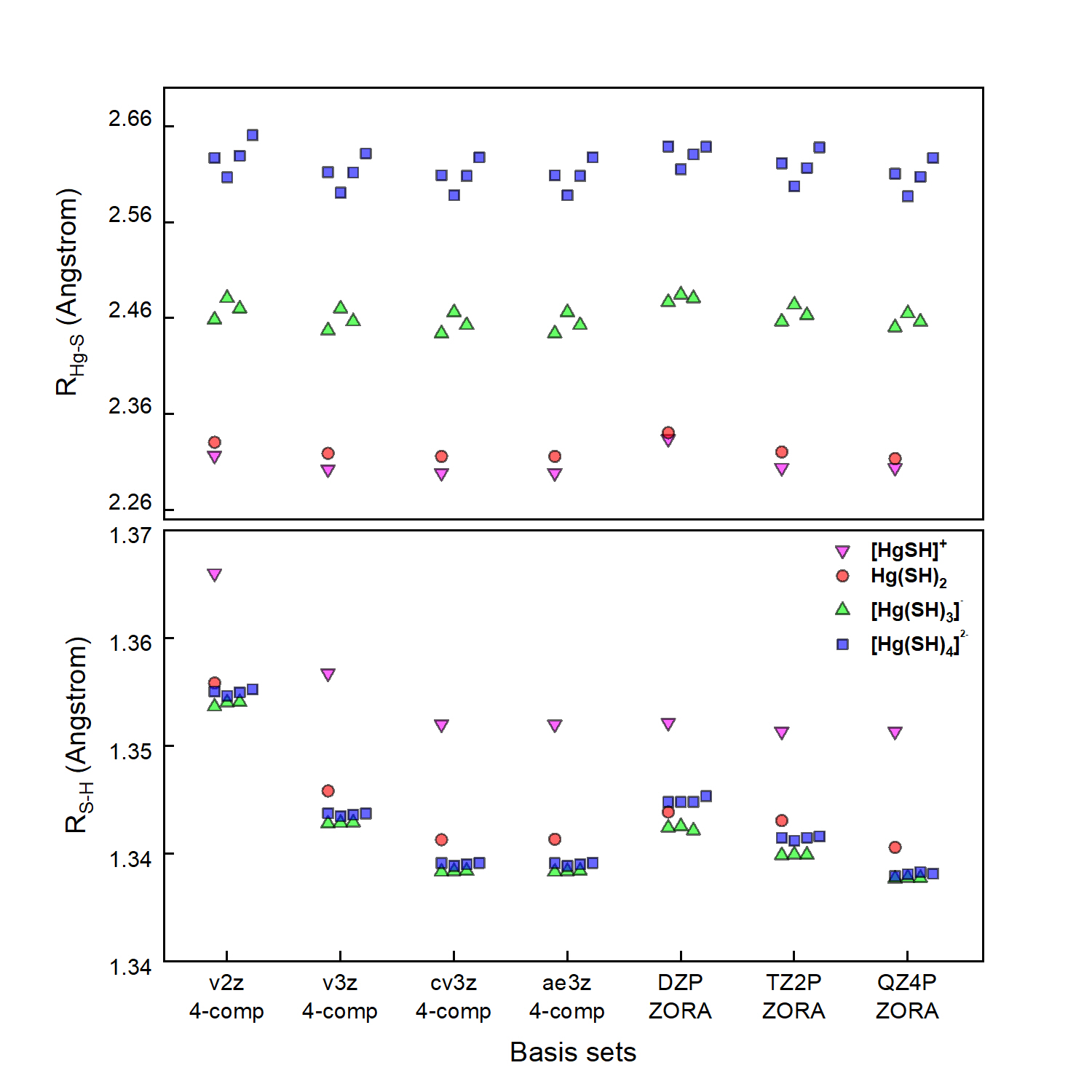}
    \caption{$R\rm_{S-Hg}$ and $R\rm_{S-H}$ bond lengths obtained by 4-component and ZORA geometry optimizations with the PBE0 functional and different basis sets 
    }
    \label{fig:BL_BS}
\end{figure}

For the study of the basis set dependence of the optimized geometries, the PBE0 functional was used in combination with different basis sets and both the 4-component and ZORA level of relativistic treatment. 
Relevant calculated geometry parameters are listed in Tables \ref{table:BL_BS} and \ref{table:BA_BS}. 
In general, the bond lengths shorten as the size of the basis sets increases while the bond angles basically stay invariant.  

\begin{table*}[]\centering
\caption{Selected bond lengths (in \AA) in [Hg(SR)$_n$]$^{2-n}$ obtained by 4-component or ZORA geometry optimizations with PBE0 functional and different basis sets}
\label{table:BL_BS}
\begin{ruledtabular}
\begin{tabular}{llccccccc}

\multirow{2}{*}{Compound} & \multirow{2}{*}{Distance} &  \multicolumn{4}{c}{4-comp} & \multicolumn{3}{c}{ZORA} \\
 & & v2z & v3z & cv3z & ae3z & DZP & TZ2P & QZ4P \\ \hline
\multirow{2}{*}{\SHs}   & $R\rm _{Hg-S}$  & 2.3166  & 2.3023  & 2.2984 & 2.2984 & 2.3338  & 2.3037 & 2.3037  \\
 & $R\rm _{S-H}$   & 1.3620  & 1.3550  & 1.3515 & 1.3515 & 1.3516  & 1.3510 & 1.3510  \\ \hline
\multirow{2}{*}{\SHd}  & $R\rm _{Hg-S}$  & 2.3304  & 2.3188  & 2.3157 & 2.3156 & 2.3403  & 2.3202 & 2.3134  \\
 & $R\rm _{S-H}$   & 1.3544  & 1.3469  & 1.3435 & 1.3435 & 1.3454  & 1.3448 & 1.3429  \\ \hline
\multirow{6}{*}{\SHt}  & $R\rm _{Hg-S1}$ & 2.4582  & 2.4470  & 2.4438 & 2.4438 & 2.4766  & 2.4561 & 2.4501  \\
 & $R\rm _{Hg-S2}$ & 2.4808  & 2.4698  & 2.4658 & 2.4658 & 2.4844  & 2.4737 & 2.4647  \\
 & $R\rm _{Hg-S3}$ & 2.4696  & 2.4563  & 2.4526 & 2.4526 & 2.4806  & 2.4627 & 2.4560  \\
 & $R\rm _{S1-H1}$ & 1.3527  & 1.3446  & 1.3412 & 1.3412 & 1.3443  & 1.3424 & 1.3408  \\
 & $R\rm _{S2-H2}$ & 1.3530  & 1.3446  & 1.3413 & 1.3413 & 1.3444  & 1.3424 & 1.3408  \\
 & $R\rm _{S3-H3}$ & 1.3530  & 1.3447  & 1.3413 & 1.3413 & 1.3441  & 1.3424 & 1.3408  \\ \hline
\multirow{8}{*}{\SHq}  & $R\rm _{Hg-S1}$ & 2.6270  & 2.6120  & 2.6088 & 2.6088 & 2.6387  & 2.6214 & 2.6072  \\
 & $R\rm _{Hg-S2}$ & 2.6068  & 2.5907  & 2.5879 & 2.5879 & 2.6152  & 2.5974 & 2.6104  \\
 & $R\rm _{Hg-S3}$ & 2.6291  & 2.6117  & 2.6082 & 2.6082 & 2.6307  & 2.6164 & 2.6270  \\
 & $R\rm _{Hg-S4}$ & 2.6507  & 2.6316  & 2.6275 & 2.6275 & 2.6386  & 2.6379 & 2.5870  \\
 & $R\rm _{S1-H1}$ & 1.3538  & 1.3453  & 1.3418 & 1.3418 & 1.3461  & 1.3436 & 1.3409  \\
 & $R\rm _{S2-H2}$ & 1.3535  & 1.3451  & 1.3417 & 1.3417 & 1.3461  & 1.3434 & 1.3411  \\
 & $R\rm _{S3-H3}$ & 1.3537  & 1.3452  & 1.3418 & 1.3418 & 1.3461  & 1.3436 & 1.3412  \\
 & $R\rm _{S4-H4}$ & 1.3539  & 1.3453  & 1.3419 & 1.3419 & 1.3465  & 1.3437 & 1.3411  \\
\end{tabular}
\end{ruledtabular}
\end{table*}

Table \ref{table:BL_BS} and Fig. \ref{fig:BL_BS} show the bond lengths calculated with different basis sets and the PBE0 functional at the 4-component and ZORA levels.
Comparing first the 4-component results with the v2z and v3z basis sets, one can see that the differences between the bond lengths calculated with the v2z and v3z basis sets are quite large (0.4$\sim$0.6$\%$ shorter for v3z).
By adding more correlating functions for the core orbitals in the cv3z and ae3z basis sets, the results can further be improved, see Fig. \ref{fig:BL_BS}. 
Similar results were yielded by ae3z and cv3z basis sets, i.e. the bond lengths are both slightly shorter than with the v3z basis set. 
However, adding correlating functions enlarges the size of the ae3z and cv3z basis sets making them basically as large as v4z, which renders these calculations computationally quite expensive.

\begin{table*}[ht!]\centering
\caption{Selected bond angles (in $^{\circ}$) in [Hg(SR)$_n$]$^{2-n}$ obtained by 4-component or ZORA geometry optimizations with PBE0 functional and different basis sets}
\label{table:BA_BS}
\centering
\begin{ruledtabular}
\begin{tabular}{llcccccccc}
\multirow{2}{*}{Compound} & \multirow{2}{*}{Angle} &  \multicolumn{4}{c}{4-comp} & \multicolumn{3}{c}{ZORA} \\
 & & v2z & v3z & cv3z & ae3z & DZP & TZ2P & QZ4P \\ \hline
 \SHs & $\rm \theta_{Hg-S-H}$ & 91.939 & 92.265 & 92.168 & {92.167} & 91.761 & {91.886} & {91.992} \\ \hline
\multirow{2}{*}{\SHd} & $\rm \theta_{Hg-S-H}$ & 93.283 & 93.884 & 93.825 & 93.825 & 93.316 & 94.055 & 94.247 \\
 & $\rm \gamma_{S1-Hg-S2}$ & 179.991 & 179.991 & 179.991 & 179.991 & 178.598 & 178.469 & 178.546 \\ \hline
\multirow{6}{*}{\SHt} & $\rm \theta_{Hg-S1-H1}$ & {94.100} & {94.913} & 94.912 & 94.912 & 93.520 & 94.584 & {94.924} \\
 & $\rm \theta_{Hg-S2-H2}$ & {94.250} & {95.080} & 95.067 & 95.067 & 95.793 & 95.032 & {95.126} \\
 & $\rm \theta_{Hg-S3-H3}$ & {94.073} & {94.815} & 94.803 & 94.803 & 95.465 & 94.707 & {94.975} \\
 & $\rm \gamma_{S1-Hg-S2}$ & {120.305} & {119.541} & {119.522} & {119.522} & 117.783 & 119.390 & 119.478 \\
 & $\rm \gamma_{S2-Hg-S3}$ & {121.591} & {121.539} & 121.530 & 121.530 & 127.025 & 120.884 & 120.525 \\
 & $\rm \gamma_{S1-Hg-S3}$ & {118.104} & {118.920} & 118.947 & 118.948 & 115.192 & 119.726 & 119.997 \\ \hline
\multirow{10}{*}{\SHq} & $\rm \theta_{Hg-S1-H1}$ & 89.464 & 91.160 & 91.219 & 91.219 & 90.404 &90.920 & 91.502 \\
 & $\rm \theta_{Hg-S2-H2}$ & 88.964 & 91.014 & 91.060 & 91.060 & 91.743 & {91.013} & {91.704} \\
 & $\rm \theta_{Hg-S3-H3}$ & 88.264 & 90.653 & 90.698 & 90.698 & 90.572 & {90.223} & {91.271} \\
 & $\rm \theta_{Hg-S4-H4}$ & 88.901 & 91.035 & 91.086 & 91.086 & 90.843 & 90.924 & 91.588 \\
 & $\rm \gamma_{S1-Hg-S2}$ & 113.025 & 111.927 & 111.800 & 111.800 & 113.273 & 111.521 & 111.844 \\
 & $\rm \gamma_{S1-Hg-S3}$ & 106.171 & 106.720 & 106.779 & 106.779 & 107.820 & 106.944 & 106.597 \\
 & $\rm \gamma_{S1-Hg-S4}$ & 116.490 & 114.968 & 114.921 & 114.920 & 107.042 & 108.585 & 108.949 \\
 & $\rm \gamma_{S2-Hg-S3}$ & 108.130 & 108.799 & 108.857 & 108.857 & 108.838 & 114.633 & 114.364 \\
 & $\rm \gamma_{S2-Hg-S4}$ & 106.205 & 106.979 & 107.026 & 107.027 & 108.793 & 107.891 & 107.346 \\
 & $\rm \gamma_{S3-Hg-S4}$ & 106.391 & 107.235 & 107.255 & 107.256 & {111.091} & 107.047 & 107.576 \\ 
\end{tabular}
\end{ruledtabular}
\end{table*}

In the ZORA calculations, similarly, the $R_{\rm{Hg-S}}$ bond lenghts obtained with the TZ2P basis set are 0.4$\sim$0.8$\%$ shorter than the DZP results, with QZ4P they are further shortened by 0.3$\sim$0.5$\%$. 
$R\rm_{S-H}$ is less sensitive to the basis sets size, from DZP to TZ2P $R\rm_{S-H}$ only decrease 0.04$\sim$0.20$\%$, while between QZ4P and TZ2P the differences are between 0.12\% and 0.18\%. 
In more detail, the TZ2P and QZ4P basis sets provide the same $R\rm_{Hg-S}$ and $R\rm_{S-H}$ in \SHs. 
However, $R\rm_{S-H}$ becomes 0.14\%, 0.12\% and 0.18\% shorter in \SHd, \SHt and \SHq on using QZ4P. 
Simultaneously, QZ4P decreases $R\rm_{Hg-S}$ by up to 0.29\%, 0.36\% and 0.54\% for \SHd, \SHt and \SHq. 

Overall, the ZORA/QZ4P results are very close to the 4-component/cv3z and 4-component/ae3z results for the bond lengths.

Comparing to average experimental Hg-S bond lengths (\SHd: 2.345$\pm$0.025 Å, \SHt: 2.446$\pm$0.015 Å, \SHq: 2.566$\pm$0.047 Å),\cite{Manceau2008} agreement with the current optimized structures is good. We note that the change in bond length through the series is slightly larger in our optimized structures, presumably because these are gas phase calculations, and including the surroundings is expected to have an effect in particular on the charged species. 

From the 4-component data for the bond angles in Table \ref{table:BA_BS}, one can observe that the bond angle of Hg-S-H in \SHs obtained with the v3z basis set is 0.35\% larger than the one from the v2z basis set.
In {\SHd} the bond angle from the v3z calculation is 0.64\% larger than result from the v2z calculation.

In the ZORA calculations, the largest $\rm{\theta_{Hg-S-H}}$ in \SHt becomes slightly smaller by 0.06\%. 
The other two $\rm{\theta_{Hg-S-H}}$ change 0.07\% and -0.06\%. The two smaller $\rm{\theta_{Hg-S-H}}$ angles tend to be become equal as the basis set size increases. A similar divergence of $R\rm_{Hg-S}$ happens in \SHq as well, while the deviation of $R\rm_{S-H}$ is below 0.002 {\AA}. 

We can conclude, that increasing the basis set size the geometries obtained with the 4-component and ZORA calculations become very similar.


\subsubsection{Dependence on XC-functionals}
\begin{figure}
    \centering
    \includegraphics[width=0.45\textwidth]{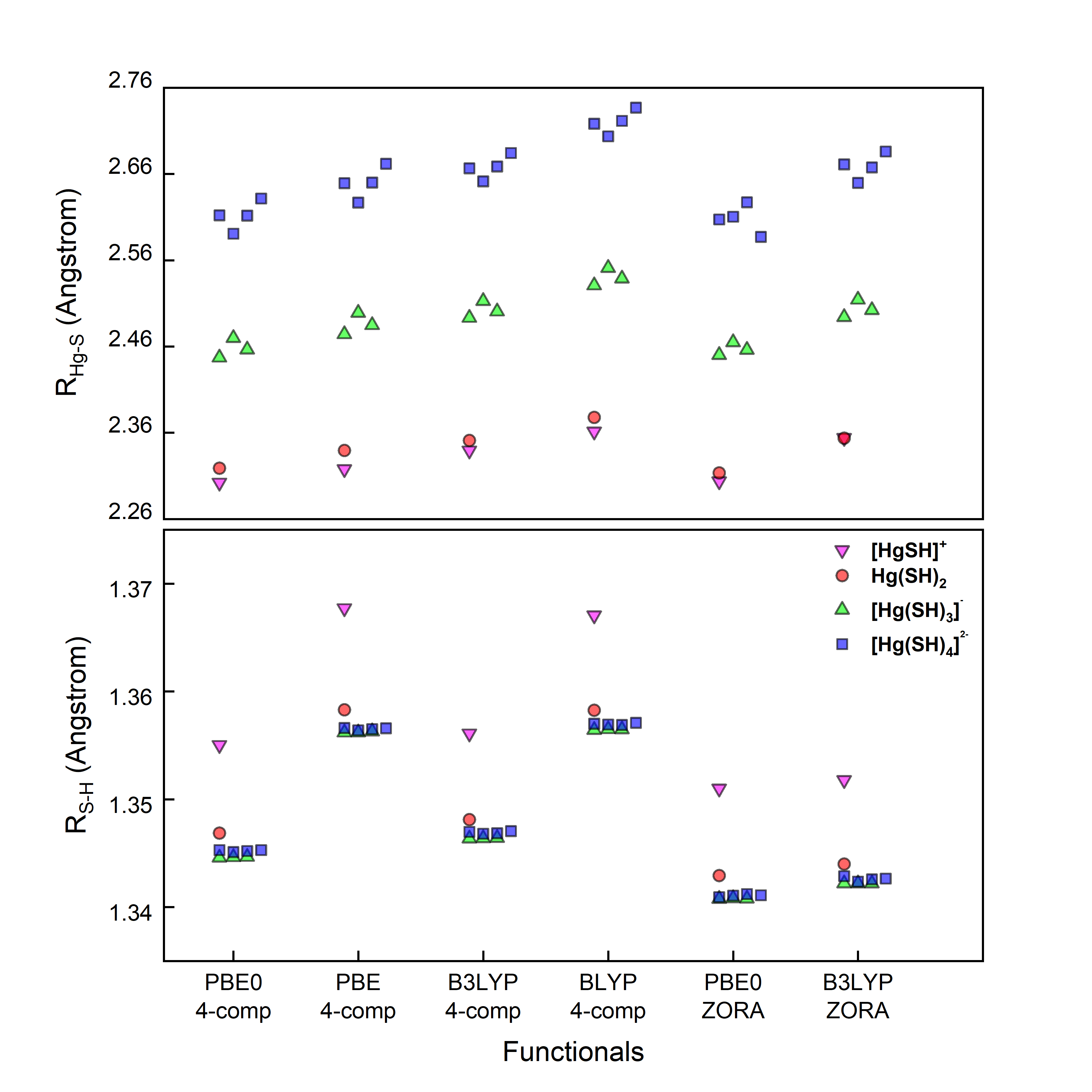}
    \caption{$R\rm_{S-Hg}$ and $R\rm_{S-H}$ bond lengths obtained by 4-component/v3z and ZORA/QZ4P geometry optimizations with different XC-functionals}
    \label{fig:SH_BL_func}
\end{figure}

The XC-functional dependence was studied using the PBE0, PBE, B3LYP and BLYP functionals together with the v3z basis set in the 4-component calculations. The ZORA calculations were performed with the PBE0 and B3LYP functionals and the QZ4P basis set. 
The optimized geometries are presented in Table \ref{table:bl_funct} and compared in Fig. \ref{fig:SH_BL_func}. 

\begin{table}[]
\centering
\caption{Bond lengths (in \AA) in [Hg(SR)$_n$]$^{2-n}$ predicted by different XC-functionals at the 4-component/v3z and ZORA/QZ4P levels.}
\label{table:bl_funct}
\begin{tabular}{llcccc|cc}
\hline
\multirow{2}{*}{Compound} & \multirow{2}{*}{Distance} &  \multicolumn{4}{c}{4-comp} & \multicolumn{2}{c}{ZORA} \\
 & & PBE0 & PBE & B3LYP & BLYP & PBE0 & B3LYP \\ \hline
\multirow{2}{*}{\SHs} & $\mathrm{R_{Hg-S}}$   & 2.3023      & 2.3179     & 2.3394& 2.3615      & 2.3037    & 2.3540     \\
                      & $\mathrm{R_{S-H}}$    & 1.3550      & 1.3677     & 1.3561& 1.3671      & 1.3510    & 1.3518     \\ \hline
\multirow{2}{*}{\SHd} & $\mathrm{R_{Hg-S}}$   & 2.3188      & 2.3394     & 2.3511& 2.3778      & 2.3134    & 2.3537     \\
                      & $\mathrm{R_{S-H}}$    & 1.3469      & 1.3583     & 1.3481& 1.3583      & 1.3429    & 1.3440     \\ \hline
\multirow{6}{*}{\SHt} & $\mathrm {R_{Hg-S1}}$ & 2.4470      & 2.4741     & 2.4932& 2.5308      & 2.4501    & 2.4941     \\
                      & $\mathrm {R_{Hg-S2}}$ & 2.4698      & 2.4990     & 2.5126& 2.5510      & 2.4647    & 2.5142     \\
                      & $\mathrm {R_{Hg-S3}}$ & 2.4563      & 2.4846     & 2.5006& 2.5386      & 2.4560    & 2.5021     \\
                      & $\mathrm R_{S1-H1}$   & 1.3446      & 1.3562     & 1.3464& 1.3564      & 1.3408    & 1.3422     \\
                      & $\mathrm {R_{S2-H2}}$ & 1.3446      & 1.3562     & 1.3464& 1.3565      & 1.3408    & 1.3422     \\
                      & $\mathrm {R_{S3-H3}}$ & 1.3447      & 1.3563     & 1.3464& 1.3565      & 1.3408    & 1.3422     \\ \hline
\multirow{8}{*}{\SHq} & $\mathrm {R_{Hg-S1}}$ & 2.6120      & 2.6493     & 2.6664& 2.7182      & 2.6072    & 2.6711     \\
                      & $\mathrm {R_{Hg-S2}}$ & 2.5907      & 2.6268     & 2.6513& 2.7036      & 2.6104    & 2.6495     \\
                      & $\mathrm {R_{Hg-S3}}$ & 2.6117      & 2.6499     & 2.6687& 2.7214      & 2.6270    & 2.6675     \\
                      & $\mathrm {R_{Hg-S4}}$ & 2.6316      & 2.6717     & 2.6842& 2.7367      & 2.5870    & 2.6859     \\
                      & $\mathrm {R_{S1-H1}}$ & 1.3453      & 1.3566     & 1.3470& 1.3570      & 1.3409    & 1.3429     \\
                      & $\mathrm {R_{S2-H2}}$ & 1.3451      & 1.3564     & 1.3468& 1.3569      & 1.3411    & 1.3424     \\
                      & $\mathrm {R_{S3-H3}}$ & 1.3452      & 1.3565     & 1.3469& 1.3569      & 1.3412    & 1.3426     \\
                      & $\mathrm {R_{S3-H3}}$ & 1.3453      & 1.3566     & 1.3471& 1.3571      & 1.3411    & 1.3427     \\ \hline
\end{tabular}
\end{table}

It can be seen from Fig. \ref{fig:SH_BL_func} that PBE0 gives the shortest bond lengths and BLYP the largest among the four XC-functionals. PBE0 and B3LYP are hybrid functionals, i.e. include a certain amount of Hartree-Fock exchange in addition to the local exchange of the GGA functionals. And PBE0 contains 5\% more exact Hartree-Fock exchange than B3LYP in the standard setup. 

The effect of changing the XC-functionals is quite different for the $\rm{R_{Hg-S}}$ and $\rm{R_{S-H}}$ bond lengths, i.e. for the bond between two light atoms and for the bond between a heavy and a light atom. As earlier observed for [Pt(CN)$_4$]$^{2-}$,\cite{o2013optimizing} the change from GGA to hybrid functionals is more important than which XC-functional is employed for the bonds between light atoms.
Looking at the bottom panel of Fig. \ref{fig:SH_BL_func}, the $\rm{R_{S-H}}$ predicted by B3LYP and PBE0 are basically the same and quite different from the PBE and B3LYP results, which are also very close to each other. The calculated $\rm{R_{S-H}}$ can thus be sorted into two groups, one for hybrid functionals and the other for GGA functionals. The key factor affecting $\rm{R_{S-H}}$ is whether the XC-functional includes exact Hartree-Fock exchange or not. 

On the other hand, $\rm{R_{Hg-S}}$ is in general more sensitive, and in particular the choice of the correlation functional exerts a larger impact. 
One can observe in the top panel of Fig. \ref{fig:SH_BL_func} that the GGA functionals always give longer $\rm{R_{Hg-S}}$ compared to their corresponding hybrid functionals. 
In \SHs and \SHd, employing the hybrid functionals shortens 
$\rm{R_{Hg-S}}$
by 0.015-0.020 {\AA} or 0.08\%. 
The difference between PBE0 and B3LYP keeps constant for all complexes, i.e. PBE0 provides shorter $\rm{R_{Hg-S}}$ (by 0.032 to 0.046 {\AA} or 1.4-1.8\%) than B3LYP, and $\rm{R_{S-H}}$ were only shortened by 0.001-0.002 {\AA} or 0.1\%.


\begin{figure}[h!]
    \centering
    \includegraphics[width=0.45\textwidth]{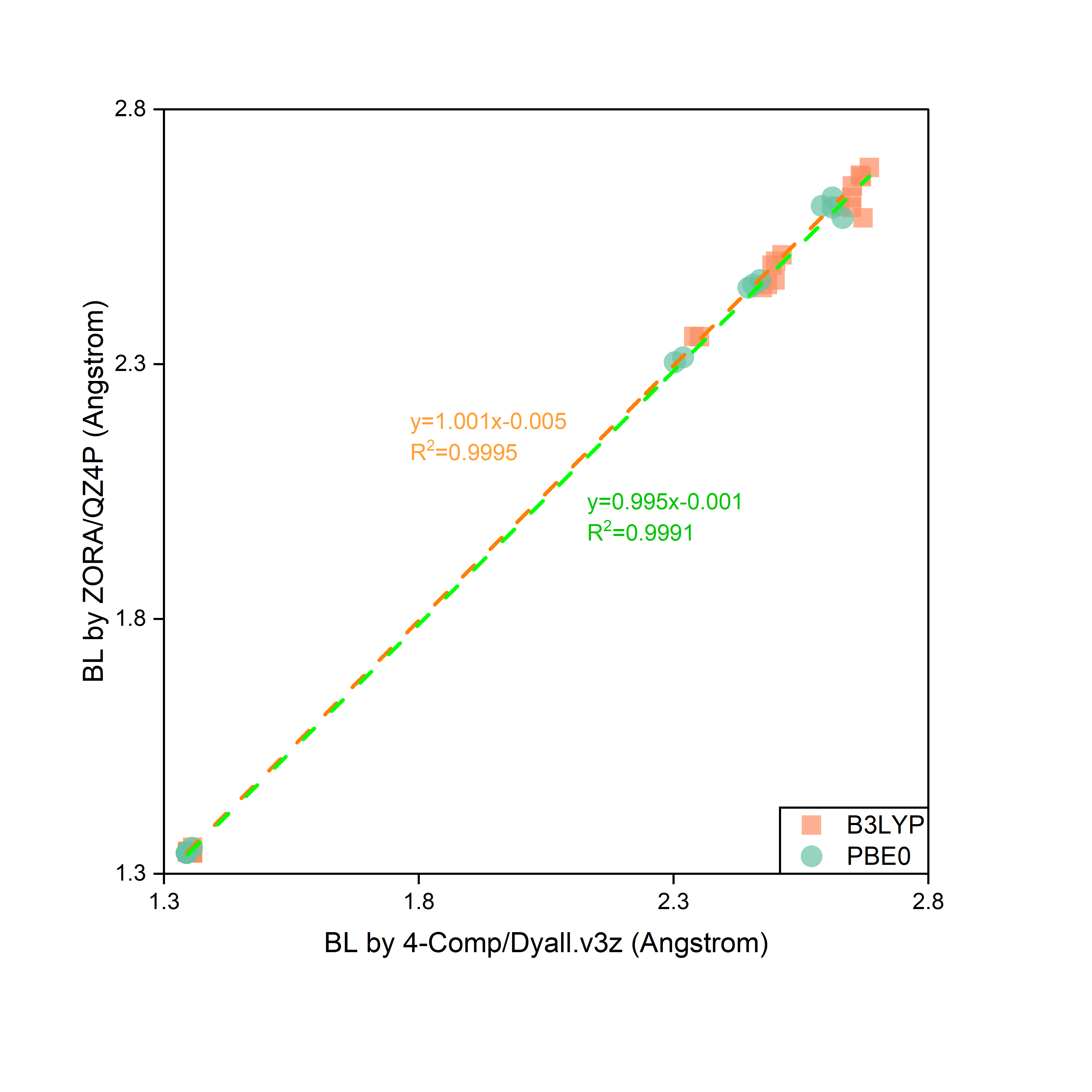}
    \caption{Correlation of $R\rm_{S-Hg}$ (Top-right points) and $R\rm_{S-H}$ (bottom-left points) bond lengths obtained by the ZORA/QZ4P and the 4-component/v3z method. ("BL" refers to bond lengths)}
    \label{fig:BL_4c_zora_funct}
\end{figure}
\subsubsection{Dependence on Relativistic Method}
Comparing finally both the $\rm{R_{S-H}}$ and $\rm{R_{Hg-S}}$ bond lengths predicted with the ZORA/QZ4P method with the corresponding 4-component/v3z results in Fig. \ref{fig:BL_4c_zora_funct} for both the PBE0 and B3LYP functional, we observe a very good correlation with an almost perfect slope and only an insignificant offset.   
Therefore, the significantly more time consuming 4-component geometry optimizations can easily be replaced by faster ZORA calculations, as we will do in the part of this study, which is concerned with the changes in the absolute shieldings to due changes in the geometry, where even the 4-component shielding calculations are carried out at ZORA optimized geometries.



\subsection{Calculation of $^{199}$Hg NMR Isotropic Shielding Constant}
Using the optimized geometries from the previous section, calculations of the isotropic $^{199}$Hg absolute shielding constant, $\sigma$, were carried out using consistently the same relativistic methods, XC-functional and basis sets as in the geometry optimization. 
It is important to note that the optimized structures used for the calculation of shielding constants therefore differ from one method to the next. 
Consequently, this is not a test of the effect of using different methods on the property (shielding) calculation alone, but rather a test of the variance of the results achieved using a given method throughout for both geometry optimization and property calculation. 
This was inspired by previous observations for calculations of spin-spin coupling constants, demonstrating that it may give better results to calculate the property for a geometry optimized structure than for an idealized "best" structure, which may not be at the potential energy minimum.\cite{spas209}

Similar to the previous section, different basis sets, XC-functional and relativistic treatment are employed. 
The same [Hg(SH)$_n$]$^{2-n}$ series of complexes will be our focus. 
Yet, the \SHs molecule will be excluded. 
It has been demonstrated\cite{samie2020coordination, jalilehvand2006mercury, carlton1990preparation, manceau2015formation} that HgL$^+$ complexes are very rare both in aqueous or in gas phase. 
Our results also show that \SHs has a very unstable electronic structure so that a slight change of geometry or calculation methods leads to a significantly different result for the $^{199}$Hg shielding. 
On the other hand, we have added \Med, which is often used as reference compound in the calculation of a chemical shift $\delta$ for $^{199}$Hg according to
\begin{equation}
    \delta=\frac{\sigma(\mathrm{Hg(CH_3)_2})-\sigma} 
                {1-\sigma(\mathrm{Hg(CH_3)_2})\times10^{-6}}
\end{equation} 
where $\sigma$(\Med) will be calculated also at optimized geometries obtained with the same basis sets and XC-functionals.

\begin{table*}[ht!]
\caption{$^{199}$Hg NMR shielding constants $\sigma$ and chemical shift $\delta$ calculated at the 4-component/PBE0 and ZORA/PBE0 levels with different basis sets. Note that the structures were geometry optimized with the same method as the property calculation, and thus the structure differs from one entry to the next.}
\label{table:nmr_shielding_bs}
\centering
\begin{tabular}{lrrrr|rrr||rrrr|rrr}
\hline
 & \multicolumn{7}{c||}{Isotropic shielding constant $\sigma$} &  \multicolumn{7}{c}{Chemical shift $\delta$}\\
 & \multicolumn{4}{c|}{4-comp} & \multicolumn{3}{c||}{ZORA}  
 & \multicolumn{4}{c|}{4-comp} & \multicolumn{3}{c}{ZORA}\\
Compound
& \multicolumn{1}{c}{v2z} 
& \multicolumn{1}{c}{v3z} 
& \multicolumn{1}{c}{cv3z} 
& \multicolumn{1}{c|}{ae3z} 
& \multicolumn{1}{c}{DZP} 
& \multicolumn{1}{c}{TZ2P} 
& \multicolumn{1}{c||}{QZ4P} 
& \multicolumn{1}{c}{v2z} 
& \multicolumn{1}{c}{v3z} 
& \multicolumn{1}{c}{cv3z} 
& \multicolumn{1}{c|}{ae3z} 
& \multicolumn{1}{c}{DZP} 
& \multicolumn{1}{c}{TZ2P} 
& \multicolumn{1}{c}{QZ4P} \\ \hline
\Med & 10482.1  & 10452.0  & 10680.6  & 10828.4  
& 9351.4   & 9205.4    & 9404.6   
&\multicolumn{1}{c}{-}&\multicolumn{1}{c}{-}&\multicolumn{1}{c}{-}&\multicolumn{1}{c|}{-}
&\multicolumn{1}{c}{-}&\multicolumn{1}{c}{-}&\multicolumn{1}{c}{-}\\
\SHd & 11417.1  & 11332.7  & 11537.5     & 11685.4     
     & 10036.2  & 9824.3    & 10209.8  & -945.0  & -890.0  & -866.2  & -866.3  
     & -691.2 & -624.6  & -812.9 \\
\SHt & 10900.7  & 10819.5  & 11030.5     & 11178.3     
     & 9480.6   & 9306.8    & 9724.6   & -423.0  & -371.4  & -353.6  & -353.7 
     & -130.5 & -102.3  & -323.1\\
\SHq & 11691.1  & 11585.7  & 11776.7  & 11924.5     
     & 10027.6  & 9956.7    & 10447.5 & -1221.9 & -1145.7 & -1107.9 & -1108.1
     &  -682.6  & -758.2  & -1052.9\\ \hline
\end{tabular}
\end{table*}

\subsubsection{Dependence on Basis Sets}
\begin{figure}[h!]
    \centering
    \includegraphics[width=0.45\textwidth]{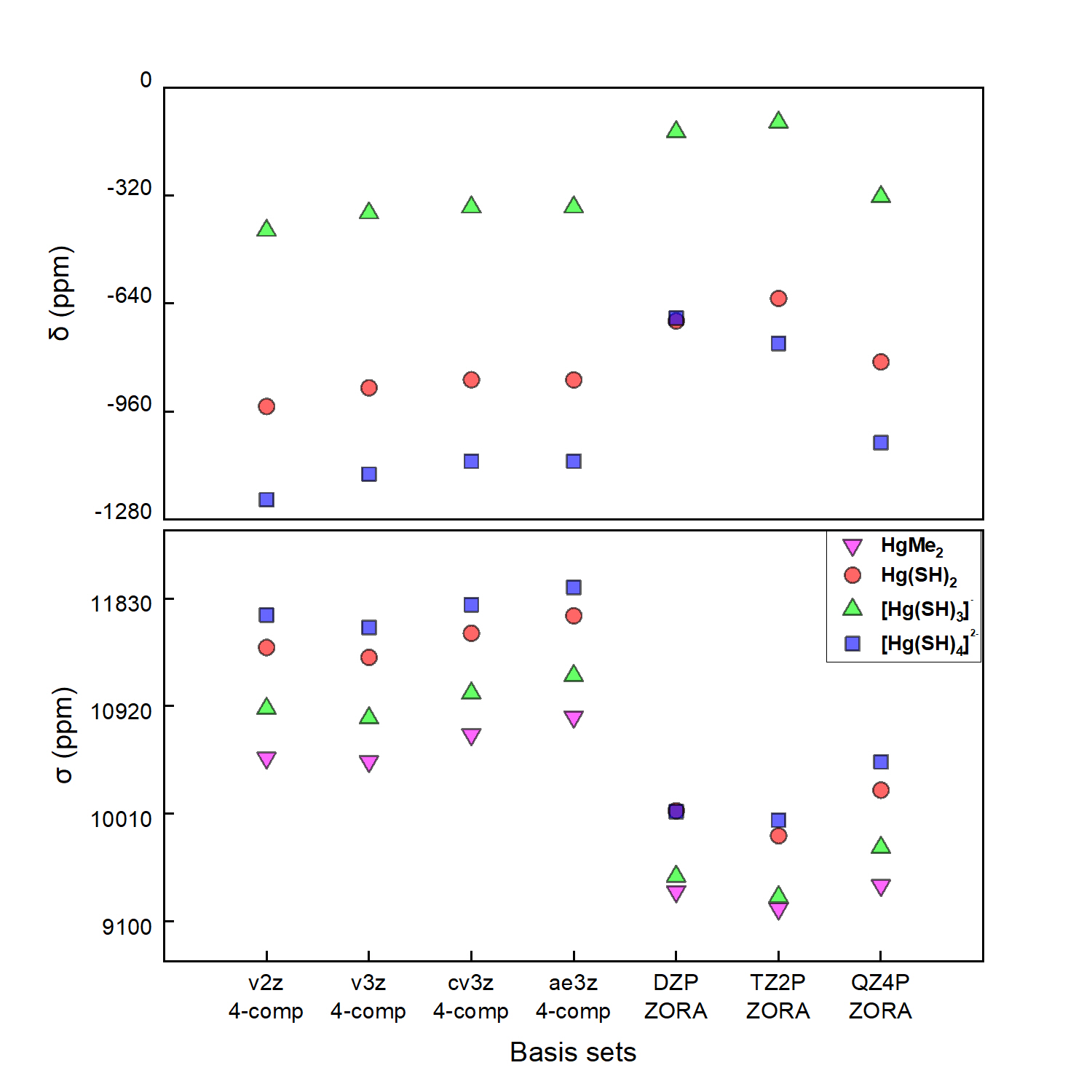}
    \caption{Isotropic shielding constant $\sigma$ and chemical shift $\delta$ of $^{199}$Hg in \Med and [Hg(SR)$_n$]$^{2-n}$ obtained with different basis sets at the 4-component and ZORA levels using the PBE0 functional. Note that the structures were geometry optimized with the same method as the property calculation, and thus the structure differs from one entry to the next.}
    \label{fig:NMRshift_bs}
\end{figure}

Analyzing the basis set dependence, the data in Table \ref{table:nmr_shielding_bs} and Fig. \ref{fig:NMRshift_bs} show that the change from the v2z to the v3z basis set is rather small, between -30 and -105 ppm for $\sigma$. 
Unfortunately, the basis set dependence is not the same for the four compounds, being smallest for \Med, larger but more or less the same for {\SHd} and {\SHt} and largest for \SHq, which implies that also the chemical shift $\delta$ exhibits a basis set dependence of 52 to 76 ppm. 
Going to the basis sets with extra correlation functions, cv3z and ae3z, a change of 190 to $\sim$230 ppm in the absolute shielding is observed for going from v3z to cv3z and from c3vz to ae3z a constant shift of 148 ppm, meaning equal for all four molecules. 
These changes are, however, very similar for the four molecules, so that the changes in chemical shift from v3z to cv3z are only between 18 ppm and 38 ppm and the further change in chemical shift to ae3z basis set is close to zero.


Turning now to the ZORA results in Fig. \ref{fig:NMRshift_bs} and Table \ref{table:nmr_shielding_bs}, we observe the same non-monotonic trend as seen before,\cite{arcisauskaite2011nuclear,spas193} the shielding is reduced by between $\sim70$ and $\sim210$ ppm on going from DZP to TZ2P but increases again from TZ2P to QZ4P by $\sim200$ to $\sim500$ ppm.
This behaviour was explained previously as the consequence of two opposing effects, i.e. an increase in the shielding constant on increasing the cardinal number from D to T and Q and a reduction of the shielding constant on adding more polarization functions,\cite{arcisauskaite2011nuclear} but in our case, the difference may in addition also originate from differences of the structures optimized with different basis sets. 
Contrary to our 4-component results, the change on going from a double to a triple $\zeta$ basis is now larger for the smaller compounds, while for the change from TZ2P to QZ4P it is \SHq and \SHd which are more affected than \SHt and \Med. 
The resulting changes in the chemical shifts are again significantly smaller in all but one case.
For the change from TZ2P to QZ4P, e.g. the change in the chemical shifts, $\sim190$ to $\sim300$ ppm, is around a factor of 2 smaller than the change in the absolute shielding.

Perhaps the most important observation is that the calculated chemical shifts, using the largest basis sets (cv3z or ae3z for the 4-component calculations and QZ4P for the ZORA calulations) give results which agree within about 55 ppm. We no that the underlying optimized structures are very similar using these basis sets. Assuming that this implies that basis set convergence has been achieved within about 55 ppm, it is very encouraging for the interpretation of experimental $^{199}$Hg chemical shifts which vary by hundreds to thousands of ppm.

Comparing with experimental $^{199}$Hg NMR data for complexes with two, three or four coordinating thiolates, the chemical shifts (-830 to -1026 ppm, -158 to -354 ppm and -485 to -793 ppm, respectively)\cite{Natan1990,Santos1991,Iranzo2007,Warner2016,Seneque2018} are of the right order of magnitude. Moreover, the calculations correctly predict the three-coordinated {\SHt} to give the least shielded Hg(II), although one should keep in mind that the shielding tensor is highly anisotropic (except for tetrahedral \SHq complexes). There are, however, also discrepancies between the calculated and experimental chemical shifts; there appears to be a trend that the charged species, in particular \SHq, are predicted to have too large shielding (and thus too low (negative) chemical shifts). This is perhaps not unexpected, because the effect of this charge may be more pronounced in the gas phase calculations than it is in solution or solid state experiments, where it is counter balanced by the surroundings.



\subsubsection{Dependence on XC-Functionals}
\begin{figure}[h]
    \centering
    \includegraphics[width=0.45\textwidth]{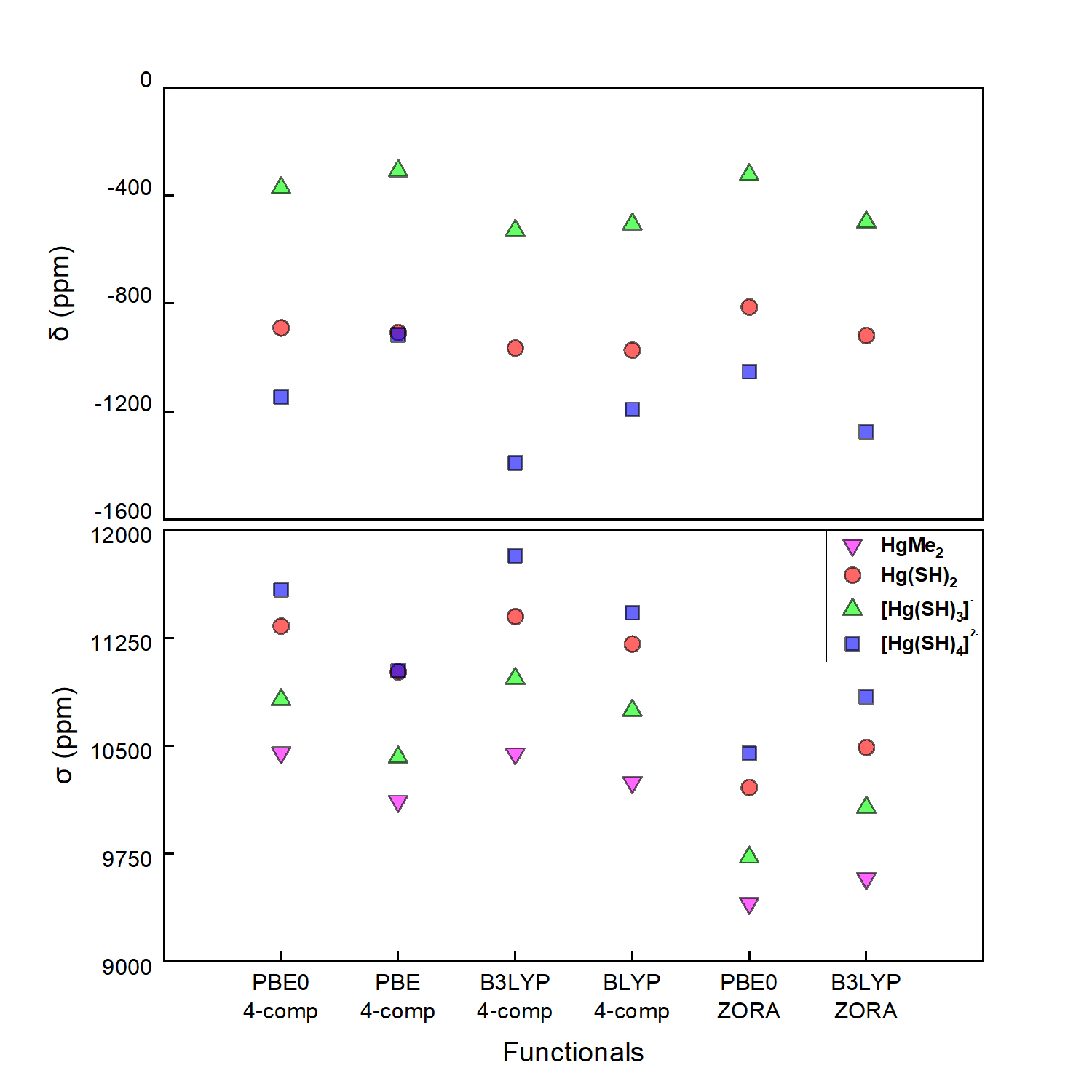}
    \caption{Isotropic shielding constant $\sigma$ and chemical shift $\delta$ of $^{199}$Hg in {\Med} and [Hg(SR)$_n$]$^{2-n}$ obtained with the different XC-functionals and the 4-component/v3z and ZORA/QZ4P methods. Note that the structures were geometry optimized with the same method as the property calculation, and thus the structure differs from one entry to the next.}
    \label{fig:NMRshift_funct}
\end{figure}

The data in Table \ref{table:nmr_shielding_funct} and Fig. \ref{fig:NMRshift_funct}, show that the shielding constant is significantly affected by the selection of XC-functional with changes up to 564 ppm. 
However, the chemical shift is not as much influenced with a maximal change of 276 ppm.
Similar to the basis set dependence of $\sigma$ in the previous section, also the dependence on the XC-functional increases with the size of the molecule. 

\begin{table}[ht!]\centering
\caption{$^{199}$Hg NMR isotropic shielding constants $\sigma$ and chemical shift $\delta$ calculated with different XC-functionals at the 4-component/v3z and ZORA/QZ4P levels. Note that the structures were geometry optimized with the same method as the property calculation, and thus the structure differs from one entry to the next.}
\label{table:nmr_shielding_funct}
\begin{tabular}{lrrrr|rr}
\hline
& \multicolumn{4}{c|}{4-comp} & \multicolumn{2}{c}{ZORA} \\
Compound 
& \multicolumn{1}{c}{PBE0} 
& \multicolumn{1}{c}{PBE} 
& \multicolumn{1}{c}{B3LYP} 
& \multicolumn{1}{c|}{BLYP} 
& \multicolumn{1}{c}{PBE0} 
& \multicolumn{1}{c}{B3LYP} 
\\ 
\hline
 & \multicolumn{6}{c}{Isotropic shielding constant $\sigma$} \\
\Med & 10452.0    & 10115.3   & 10443.7     & 10245.7    & 9404.6   & 9578.5  \\
\SHd & 11332.7    & 11013.6   & 11398.6     & 11208.3    & 10209.8  & 10488.1 \\
\SHt & 10819.5    & 10419.8   & 10967.2     & 10745.5    & 9724.6   & 10072.0 \\
\SHq & 11585.7    & 11021.8   & 11820.1     & 11425.5    & 10447.5  & 10841.4 \\ \hline
 & \multicolumn{6}{c}{Chemical shift $\delta$}\\
\SHd      & -890.0     & -907.5     & -965.3     & -972.5      & -812.9    & -918.4 \\
\SHt      & -371.4     & -307.6     & -529.0     & -505.0      & -323.1    & -498.2 \\
\SHq      & -1145.7    & -915.8     & -1391.0    & -1192.0     & -1052.9   & -1275.1 \\ \hline
\end{tabular}
\end{table}

Looking first at the effect on $\sigma$ of using a hybrid functional, i.e. comparing PBE0 to PBE and B3LYP to BLYP, the hybrid functionals give consistently larger shielding constants. 
The results with PBE0 are $\sim350$ to $\sim550$ ppm larger than the PBE results and the B3LYP results are $\sim200$ to $\sim400$ ppm larger than the BLYP values. 
The difference between $\sigma$ predicted by PBE0 and by B3LYP varies equally with the molecules but is in the 4-component calculations somewhat smaller, i.e. up to $\sim250$ ppm, while it is with $\sim150$ to $\sim400$ ppm larger in the ZORA calculations.
Generally, the isotropic shielding increases from PBE over BLYP over PBE0 to B3LYP. Correspondingly, the absolute values of the chemical shifts decrease from B3LYP over BLYP to PBE0 and PBE.

Despite the changes in the absolute values, a pattern can be seen from Fig. \ref{fig:NMRshift_funct} with respect to the relative values of both the shielding constants and chemical shifts of the different molecules that are relatively consistent with 
$\sigma$(\Med) < $\sigma$(\SHt) < $\sigma$(\SHd) < $\sigma$(\SHq), while  $\sigma$(\SHs) – $\sigma$(\SHt) $\sim400$ to 600 ppm and  $\sigma$(\SHq)–$\sigma$(\SHt) $\sim600$ to 900 ppm. 
Furthermore, the calculated chemical shifts follow a similar pattern that $\delta$(\SHd)–$\delta$(\SHt) $\sim500$ppm and $\delta$(\SHq)–$\delta$(\SHt) $\sim600$ to 800 ppm.

The main conclusion is that the variation of calculated shielding constants and chemical shifts using the different functionals (for both geometry optimization and property calculation) is on the order of 100-200 ppm. The hybrid functionals give larger shielding constants, and this may relate to the property calculation, and not only the differences in optimized structures, because the Hg-S bond lengths change relatively little, see Table \ref{table:BL_BS}.


\subsubsection{Dependence on Relativistic Method}
\begin{figure}[h!]
    \centering
    \includegraphics[width=0.45\textwidth]{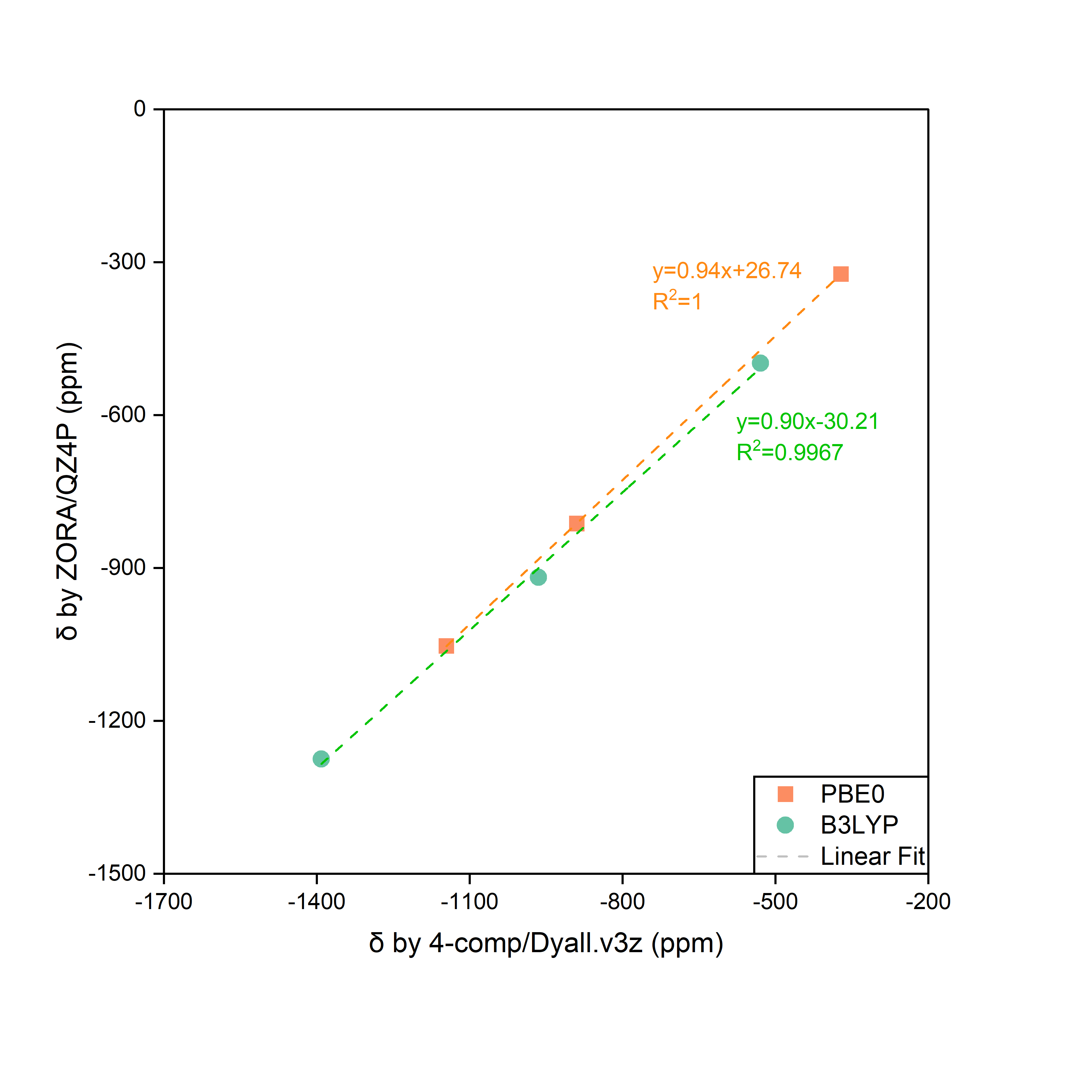}
    \caption{Chemical shift $\delta$ of mercury in [Hg(SR)$_n$]$^{2-n}$. Comparison between ZORA and 4-component calculations.}
    \label{fig:NMR_4c_zora_funct}
\end{figure}

In this section we will shortly analyze the results for the $^{199}$Hg isotropic shielding constants $\sigma$ and chemical shift $\delta$ from the previous section with respect to how they depend on the method for treating the relativistic effects, i.e. 4-component fully relativistic or approximate two-component ZORA calculations.
It is well known (e.g. Ref. \onlinecite{arcisauskaite2011nuclear}), that two-component ZORA calculations are not able to reproduce the results from corresponding 4-component calculations for the absolute shielding constants but very well reproduce the trends and thus also the chemical shifts.
Plotting the chemical shifts calculated by ZORA against the 4-component results in 
Fig. \ref{fig:NMR_4c_zora_funct} for both hybrid functionals nicely confirms this again. 
For both hybrid functionals, the three chemical shifts lie almost perfectly on a line. 
For PBE0 the fit is closer to the ideal line, with a slightly lower interception, a slope closer to 1.000 than for B3LYP. 
Nevertheless, for both XC-functionals the absolute difference between the two relativistic methods is predictable and systematic.

\begin{figure}
    \centering
    \includegraphics[width=0.45\textwidth]{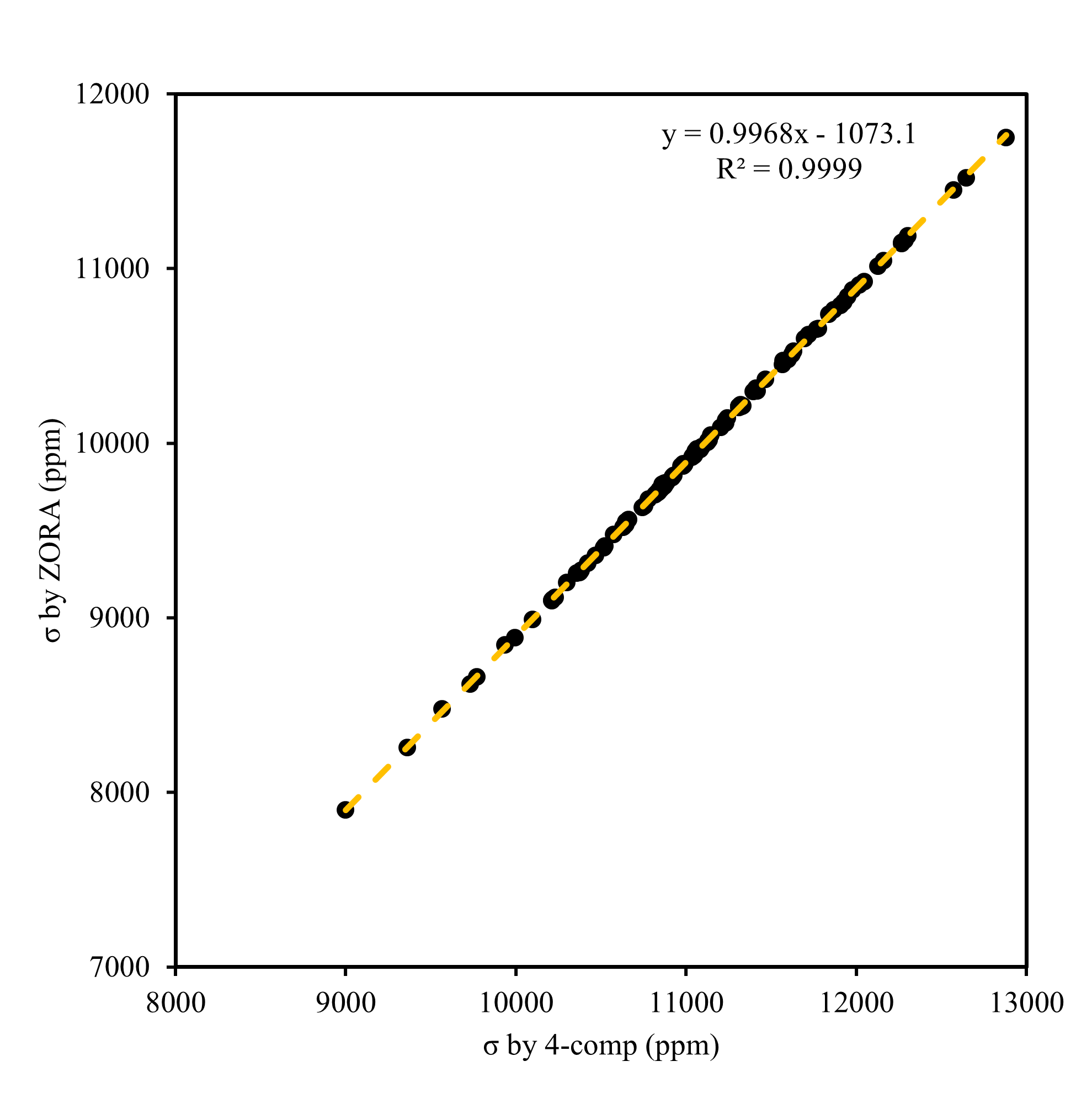}
    \caption{NMR shielding constant calculated by ZORA/PBE0/QZ4P vs. 4Comp/PBE0/v3z}
    \label{fig:zoravs4comp}
\end{figure}
As we in the following section will discuss with both 4-component and ZORA calculations, how the isotropic shielding constants vary with changes in the geometry for a larger set of molecules, i.e. for Hg(SR)$_2$(R=H, methyl, ethyl and phenyl) and \SHt, we have produced both 4-component and ZORA results for $\sigma$ for many modified molecular geometries. 
In Fig. \ref{fig:zoravs4comp} we correlate thus over 150 data points in different complexes with various geometries to determine the correlation between the ZORA and four-component results for the isotropic shielding constant. 
It is highly obvious that the difference in shielding constant calculated with ZORA and 4-component is a constant $\sim$1073 ppm. 
The data can be perfectly fitted by a linear regression with the coefficient of determination being $R=0.9999$ and a slope of 0.9968. This proofs once again that ZORA has the possibility to almost perfectly reproduce 4-component results for chemical shifts.



\subsection{Geometry Dependence of the $^{199}$Hg Isotropic Shielding Constant}
All the calculations of absolute shieldings or chemical shifts in the previous section were carried out at geometries optimized at the same level of theory.
This will often also be the standard approach for interpretating experimental NMR spectra. Nevertheless, further information on the structure of Hg-complexes can be extracted, if one knows how the $^{199}$Hg chemical shift or isotropic shielding constant varies with changes in the geometry, i.e. with deviations from optimizied geometries. In the following we will try to answer this question.

In order to see how the $^{199}$Hg shielding constant is affected by changes in the molecular geometry, we calculated it for a series of modified molecular geometries focusing in particular on changes in the bond lengths or in the bond angles. 
The modified molecular geometries were generated by first choosing a fixed value for a particular bond length or bond angle and then re-optimizing the remaining bond lengths and angles.
At these partially optimized geometries the $^{199}$Hg shielding constant was calculated.
The change of the shielding constant and optimized bond lengths will be plotted against the modification we applied. 
A normalized percentage will be used as unit for all data in this section, i.e. for the shielding, 
\begin{equation}
    \Delta\sigma=\frac{\sigma_{\mathrm {M}}-\sigma_{\mathrm {orig}}}{\sigma_{\mathrm {orig}}}\times100\%
\end{equation}
where $\Delta\sigma$ is the percentage change of the shielding constant, $\sigma_{\mathrm {M}}$ is the shielding constant calculated at the modified geometry and $\sigma_{\mathrm {orig}}$ is the value at the original fully optimized geometry.
For geometric parameters, taking the sulfur-mercury bond length as an example, we look at the percentage change in the modified bond
\begin{equation}
    \Delta R_{\mathrm{(Hg-S),M}}=\frac{R_{\mathrm {(Hg-S),M}}-R_{\mathrm {(Hg-S),orig}}}{R_{\mathrm {(Hg-S),orig}}}\times100\%
\end{equation}
and in the other, re-optimized bonds
\begin{equation}
    \Delta R_{\mathrm{(Hg-S),Opt.}}=\frac{R_{\mathrm {(Hg-S),Opt.}}-R_{\mathrm {(Hg-S),orig}}}{R_{\mathrm {(Hg-S),orig}}}\times100\%
\end{equation}
To what extend a geometric modification affects the results can thus be seen intuitively in the following figures. For an instance, modifying a bond length by $x\%$ will cause the shielding constant change by $y\%$.

We firstly looked into the Hg(SR)$_{2}$ complexes with side groups "R" being hydrogen, methyl, ethyl, phenyl, cysteine. 
The Hg-S bond(s) or S-Hg-S angle(s) were modified by $\pm10\%$ based on the optimized geometry. 
Later, constrained geometry optimizations were carried out with the modified bond(s) or angles fixed. 
At these geometries, $^{199}$Hg shielding constants were calculated.

In the following we will only present the results of the ZORA/PBE0/QZ4P calculations, as the corresponding 4-component/PBE0/v3z calculations of the shielding constants lead to exact the same conclusions. Figures of the absolute changes in the shielding constant on changes in bond lengths or angles (Figures S1 to S8 or Tables S1 to S8 in the supplementary material) show that the ZORA and 4-component results for the changes in $\sigma$ are basically identical.

\begin{figure}[h]
    \centering
    \includegraphics[width=0.45\textwidth]{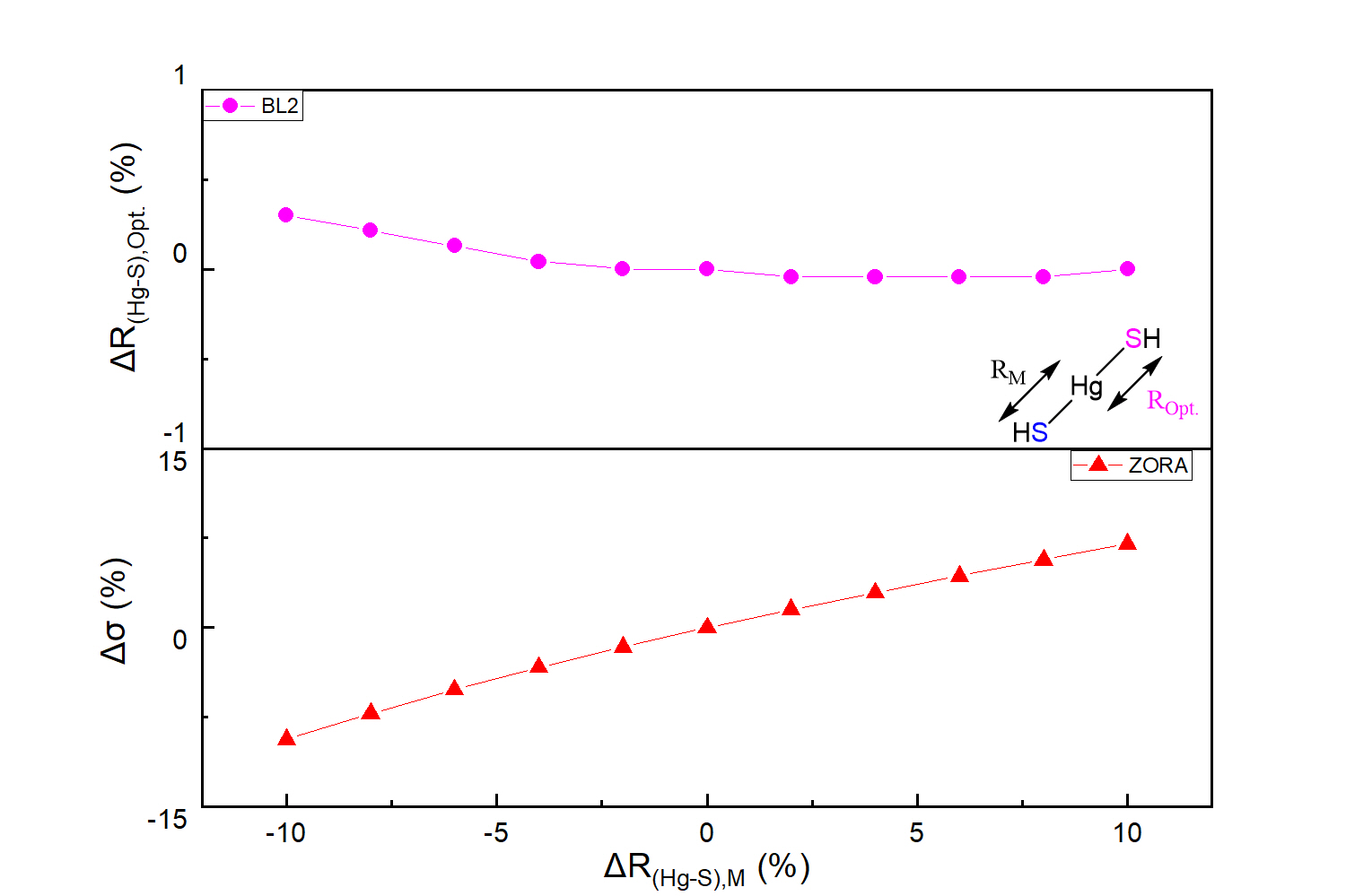}
    \caption{\SHd: $\sigma$($^{199}$Hg) isotropic shielding constant variation vs. changes in the Hg-S bond length calculated at the ZORA/PBE0/QZ4P level of theory.}
    \label{fig:HgSH2_BL1}
\end{figure}

\subsubsection{Hg(SH)$_{2}$}
\begin{figure}[h]
    \centering
    \includegraphics[width=0.45\textwidth]{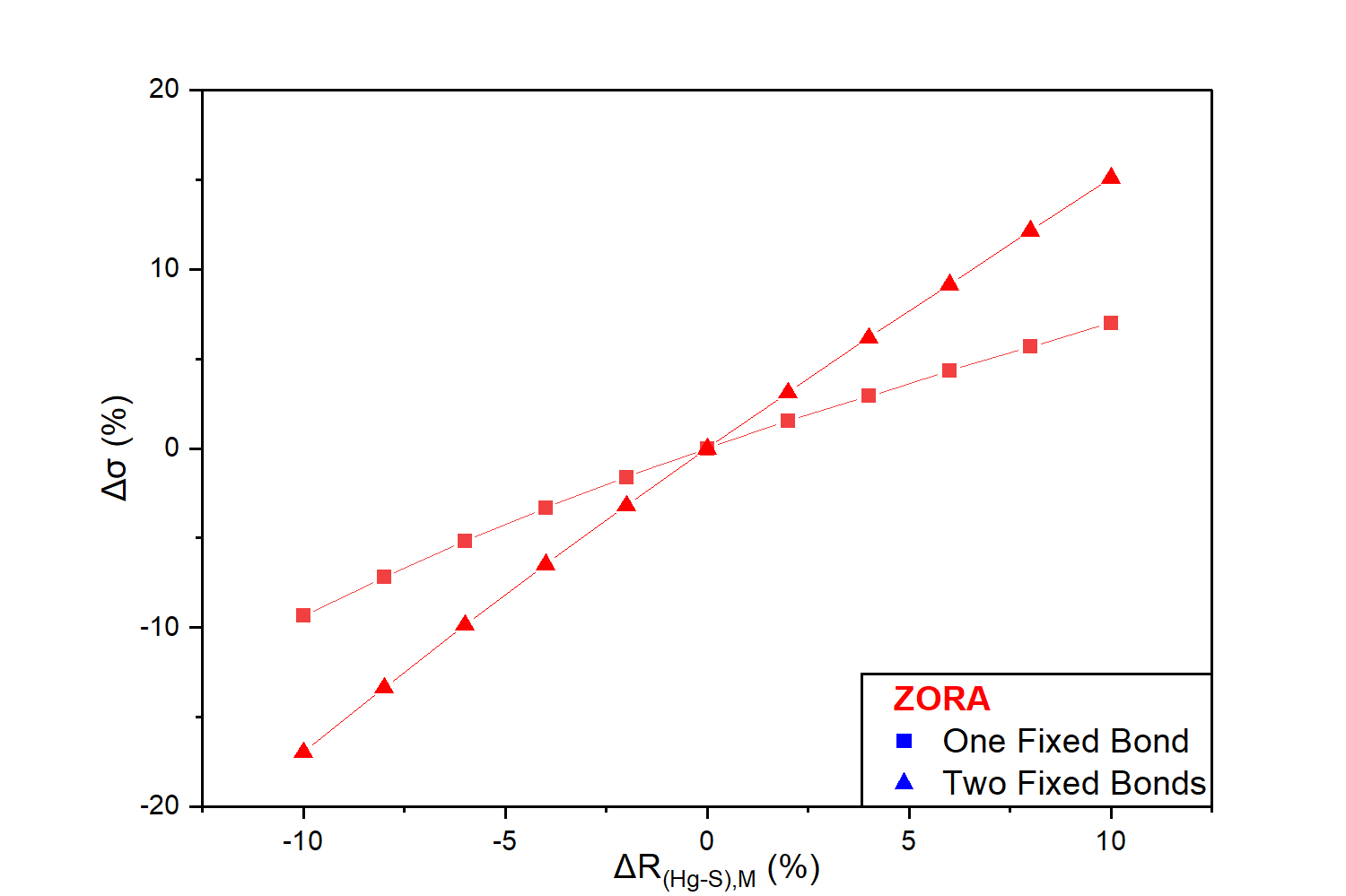}
    \caption{\SHd: $\sigma$($^{199}$Hg) isotropic shielding constant variation vs. changes in one or both Hg-S bond lengths, calculated at the ZORA/PBE0/QZ4P level of theory.}
    \label{fig:HgSH2_BL_zora}
\end{figure}

Starting with the simplest case, the {\SHd} molecule, the partial geometry optimization was performed at DFT/PBE0 level with ZORA relativistic treatment and the QZ4P basis set for all atoms. The $^{199}$Hg shielding constants were afterwards calculated with the ZORA method at the PBE0/QZ4P level and with 4-component calculations at the PBE0/v3z level. 
With one Hg-S bond fixed at a certain value, the variation of $^{199}$Hg shielding constant $\Delta\sigma$ and of the free bond length $\Delta \mathrm{R_{Opt.}}$ were plotted vs. the constraint condition $\Delta \mathrm{R_{M}}$ in Fig. \ref{fig:HgSH2_BL1}. 
The results of the constrained geometry optimization show that the free bond is not significantly effected by the changes in the fixed bond length. 
The maximal change in the free Hg-S bond, $\Delta \mathrm{R_{Opt.}}$, is only 0.3\% for the maximal change in the fixed bond length, $\pm$10\%.
The general trend is that it sort of compensates for the changes in fixed bond length, i.e. it becomes slightly larger, when the fixed bond is shortened. 
The shielding constant varies from -9.3\% to 7.0\% with the modification in the fixed bond ranging from -10\% to 10\%. 
However, when both Hg-S bonds are modified simultaneously with the same amplitude by up to $\pm10\%$, see Fig. \ref{fig:HgSH2_BL_zora}, the shielding constant changes more, i.e. up to $\pm15\%$.
In both cases, $\sigma$($^{199}$Hg) increases, when the Hg-S bond(s) are extended. Analyzing the contributions to the shielding constant one finds that it is the paramagnetic contribution, which is responsible for this change. 

\begin{figure}[h]
    \centering
    \includegraphics[width=0.45\textwidth]{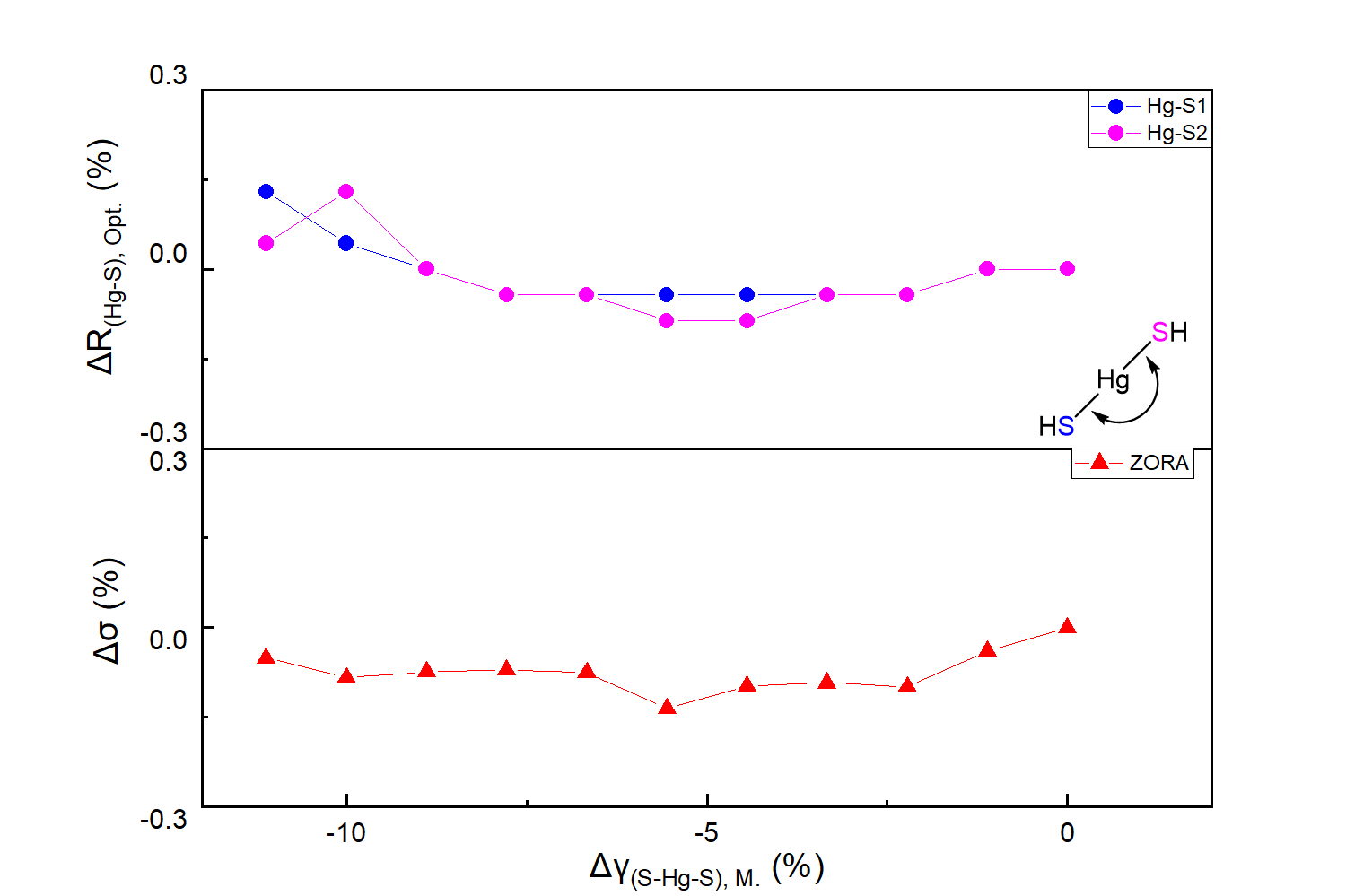}
    \caption{\SHd: $\sigma$($^{199}$Hg) isotropic shielding constant variation vs. changes in the S-Hg-S bond angle, calculated at the ZORA/PBE0/QZ4P level of theory.
}
    \label{fig:HgSH2_ba}
\end{figure}
In Fig. \ref{fig:HgSH2_ba} (Please note that the scale of y-axis is much smaller than in Fig. \ref{fig:HgSH2_BL1}) we show how $\sigma$($^{199}$Hg) changes when the S-Hg-S bond angle $\Delta \mathrm{\gamma_{M}}$ is modified by up to $-20^{\circ}$ with the two Hg-S bonds being free to adjust. 
The results in Fig. \ref{fig:HgSH2_ba} show that the modification of S-Hg-S bond angle basically does not effect neither the isotropic shielding nor the bond lengths.
The variation of bond length is lower than the convergence criteria ($10^{-4}$) employed for the optimization of the bond lengths.


\subsubsection{Hg(SR)$_{2}$}
\begin{figure}
\centering
\begin{subfigure}[b]{0.2\textwidth}
  \centering
  \includegraphics[width=1\textwidth]{HgSMe2.jpg}
  \caption{\label{HgSMe2}\SMed} 
\end{subfigure}
\hfill
\begin{subfigure}[b]{0.2\textwidth}
  \centering
  \includegraphics[width=1\textwidth]{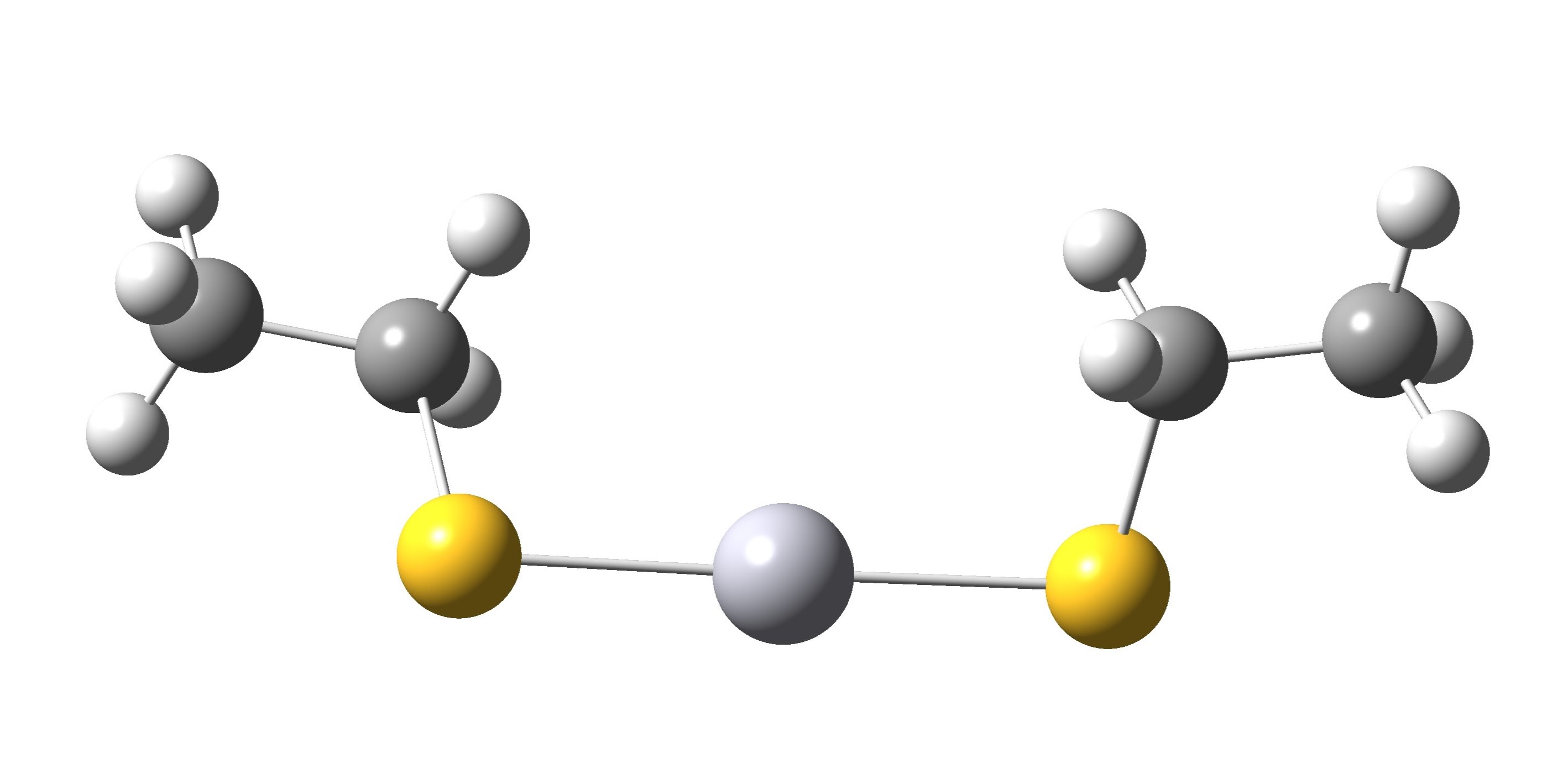}
  \caption{\label{HgSEt2}\SEtd} 
\end{subfigure}
\hfill
\begin{subfigure}[b]{0.2\textwidth}
  \centering
  \includegraphics[width=1\textwidth]{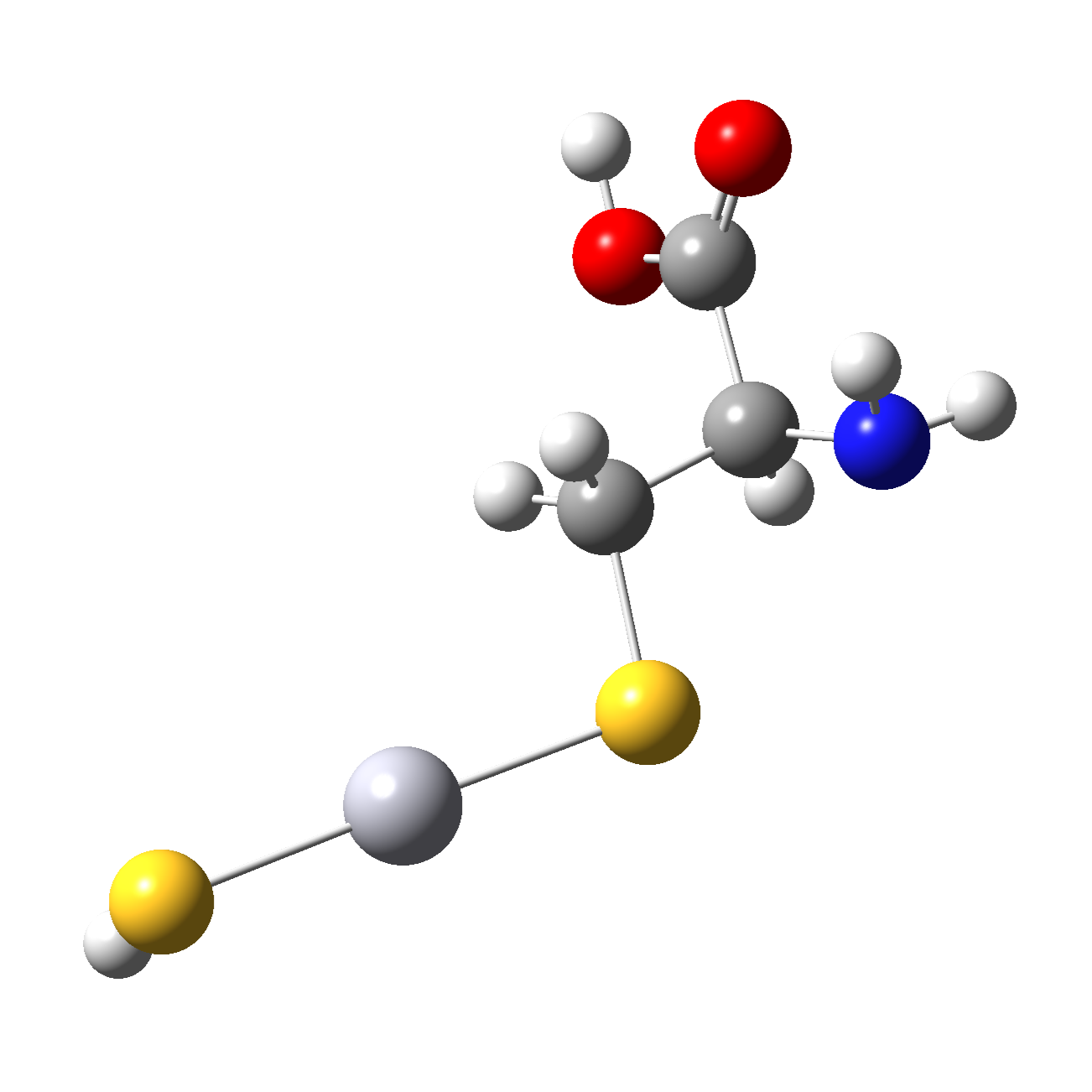}
  \caption{\label{HgSHCys}\asymcys} 
\end{subfigure}
\hfill
\begin{subfigure}[b]{0.2\textwidth}
  \centering
  \includegraphics[width=1\textwidth]{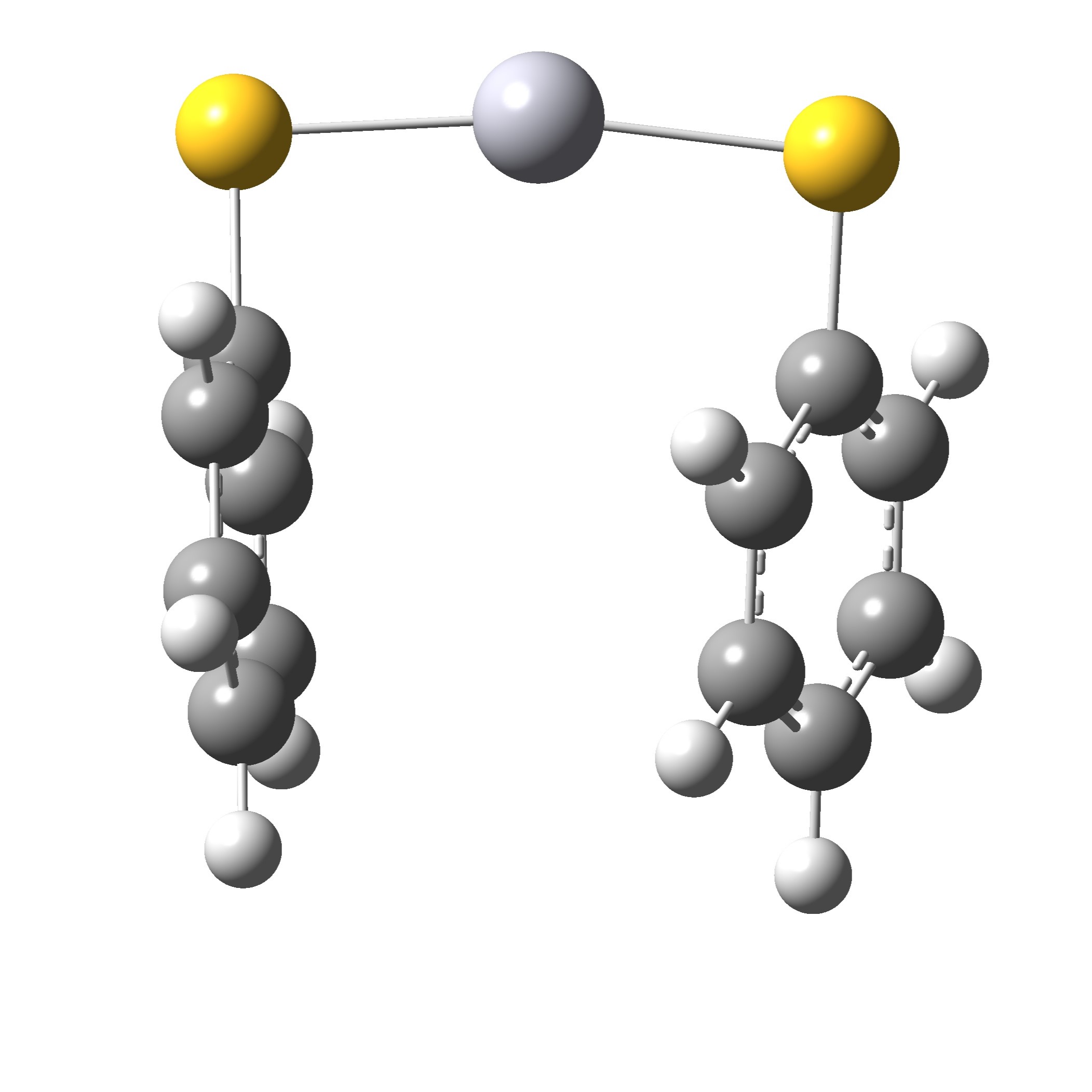}
  \caption{\label{HgSPh2}\SPhd} 
\end{subfigure}
\hfill
\begin{subfigure}[b]{0.3\textwidth}
  \centering
  \includegraphics[width=1\textwidth]{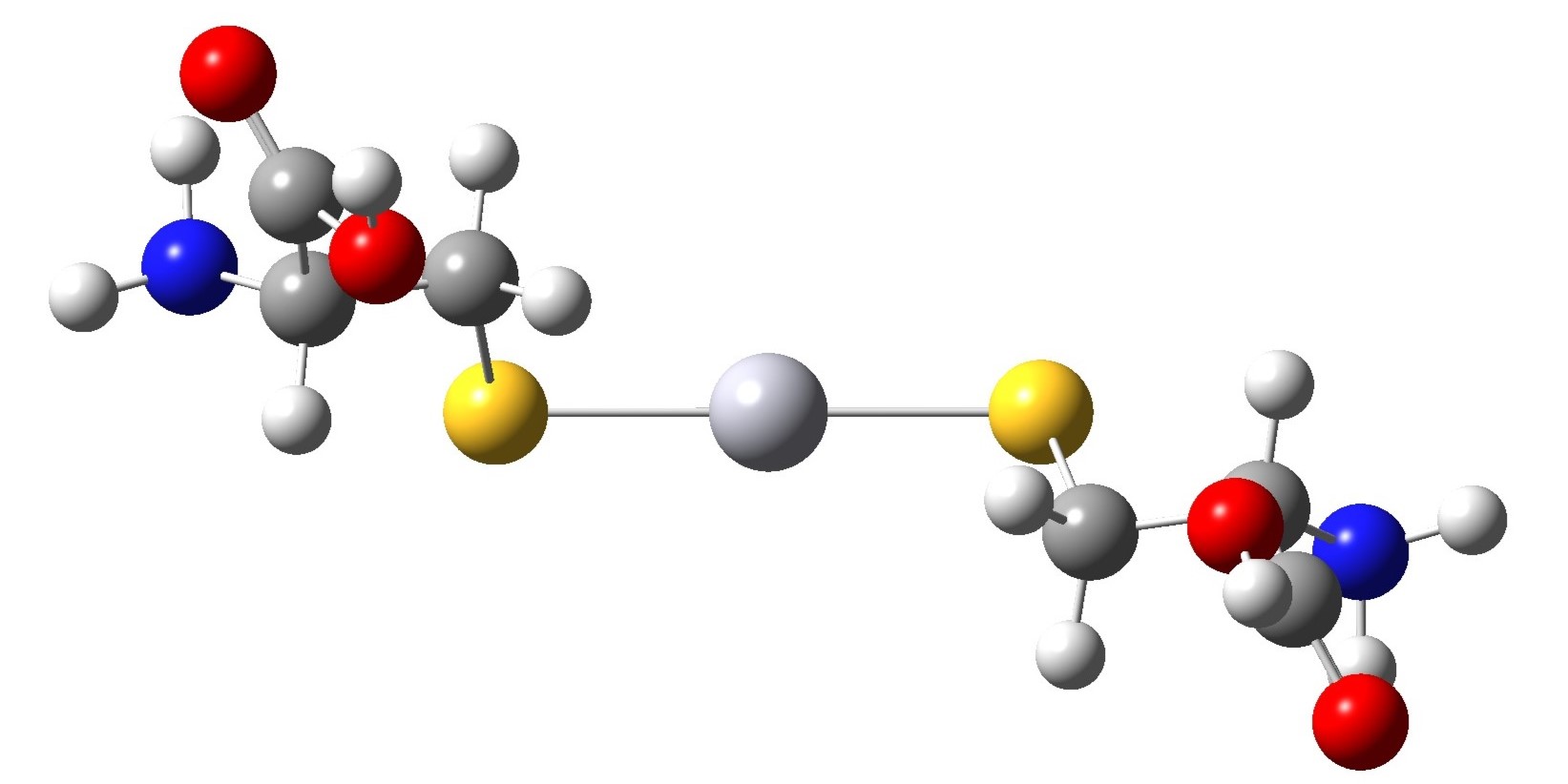}
  \caption{\label{Hgcys2}\Cysd} 
\end{subfigure}
\caption{Geometries of [Hg(SR)$_n$]$^{2-n}$, optimized at ZORA/PBE0/QZ4P level of theory.}
\label{fig:molecular.geometries.2}
\end{figure}
In order to investigate how much the changes in $\sigma$($^{199}$Hg) due to the variation in the Hg-S bond lengths are influenced by which thiolate ligands are bound to Hg, we have carried out the same type of investigation also for more complicated cases Hg(SR)$_2$ with the side groups R=methyl, ethyl, phenyl, cysteine, see Fig. \ref{fig:molecular.geometries.2}. 
Constrained geometry optimization and shielding calculations were performed at the PBE0 level with the ZORA relativistic method like for \SHd. 
Yet, the QZ4P basis set was only employed for the mercury and sulfur atoms, while TZ2P was used on carbon atoms and DZP on other atoms. 

\begin{figure}[h]
    \centering
    \includegraphics[width=0.45\textwidth]{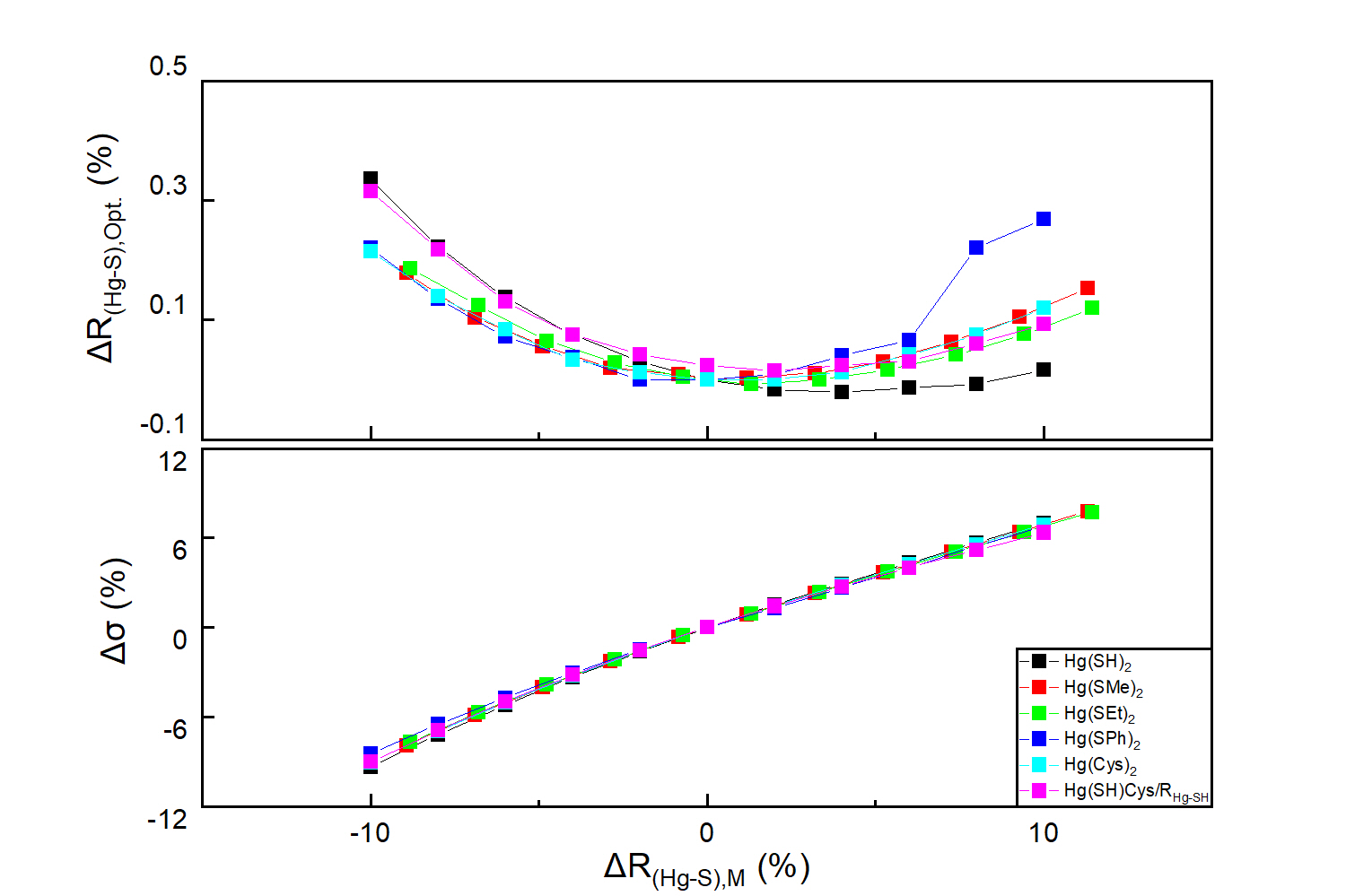}
    \caption{Hg(SR)$_2$: $\sigma$($^{199}$Hg) isotropic shielding constant variation vs. changes in one Hg-S bond length, calculated at the ZORA/PBE0 level of theory.
    }
    \label{fig:HgSR2_1bl}
\end{figure}

Fig. \ref{fig:HgSR2_1bl} shows the changes in $\sigma$($^{199}$Hg) on changing one Hg-S bond for \SMed, \SEtd, \SPhd, \Cysd, {\asymcys} together with \SHd. 
With modification of one bond length, the changes of the free bond and shielding constant are highly similar for all Hg(SR)$_2$ complexes. 
Lines from different complexes basically overlap each other. 
By modifying one Hg-S bond $\Delta \mathrm{R_{M}}$ from -10\% to 10\%, the free bonds $\Delta \mathrm{R_{Opt.}}$ were not affected significantly (maximum $\sim0.3\%$) while the shielding constants changed by $-8.4\%\sim7.7\%$ 
In the asymmetric case, {\asymcys} has two non-equivalent Hg-S bonds. 
They were investigated by separately modifying only one of them at a time. 
Thus {\asymcys} gives rise to two lines in Fig. \ref{fig:HgSR2_1bl}. 
Based on Fig. \ref{fig:HgSR2_1bl}, we can conclude that for the percentage changes in $\sigma$($^{199}$Hg) due to the variation in the Hg-S bond it is irrelevant with side groups are bound to S, because the changes are mostly influenced by Hg-S bond length. 
Furthermore, the results for {\asymcys} suggest that the percentage changes in $\sigma$($^{199}$Hg) due to changes in the two non-equivalent Hg-S bonds in Hg(SR)$_2$ are equal.

\begin{figure}[h]
    \centering
    \includegraphics[width=0.45\textwidth]{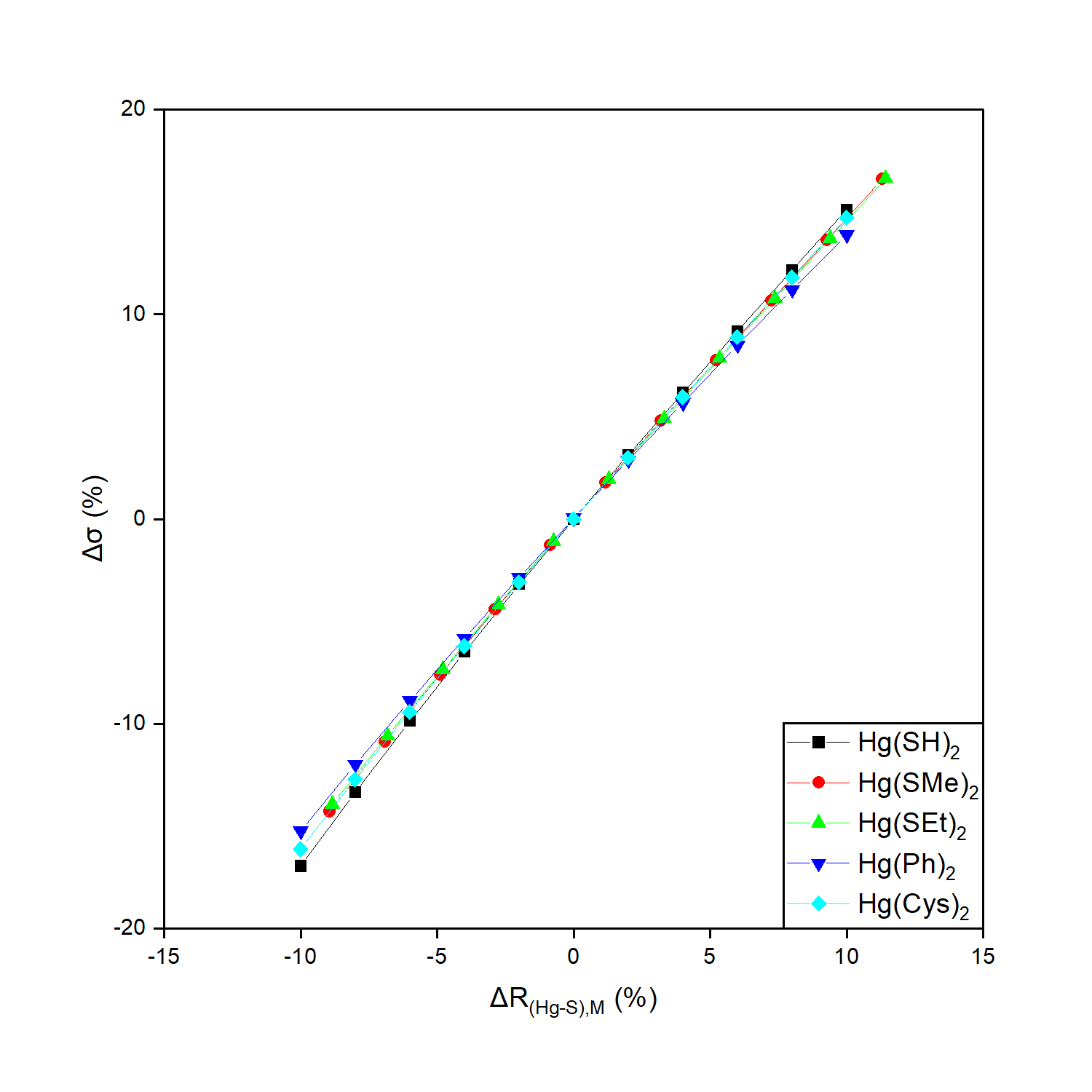}
    \caption{Hg(SR)$_2$: $\sigma$($^{199}$Hg) isotropic shielding constant variation vs. changes in both Hg-S bond lengths, calculated at the ZORA/PBE0 level of theory.}
    \label{fig:HgSR2_2bl}
\end{figure}

Fig. \ref{fig:HgSR2_2bl} shows the change of $^{199}$Hg shielding constant, when both Hg-S bonds are modified  simultaneously. 
And again $\sigma$($^{199}$Hg) increases with increasing Hg-S bond length.
The percentage changes in $\sigma$($^{199}$Hg) are larger than when only one bond was modified, as we had already seen for {\SHd} in Fig. \ref{fig:HgSH2_BL_zora}.
The overlapping lines by different compounds indicate that, in percentage unit, side groups do not affect the changes in $^{199}$Hg shielding caused by changing the Hg-S bonds lengths. 

\begin{figure}[h]
    \centering
    \includegraphics[width=0.45\textwidth]{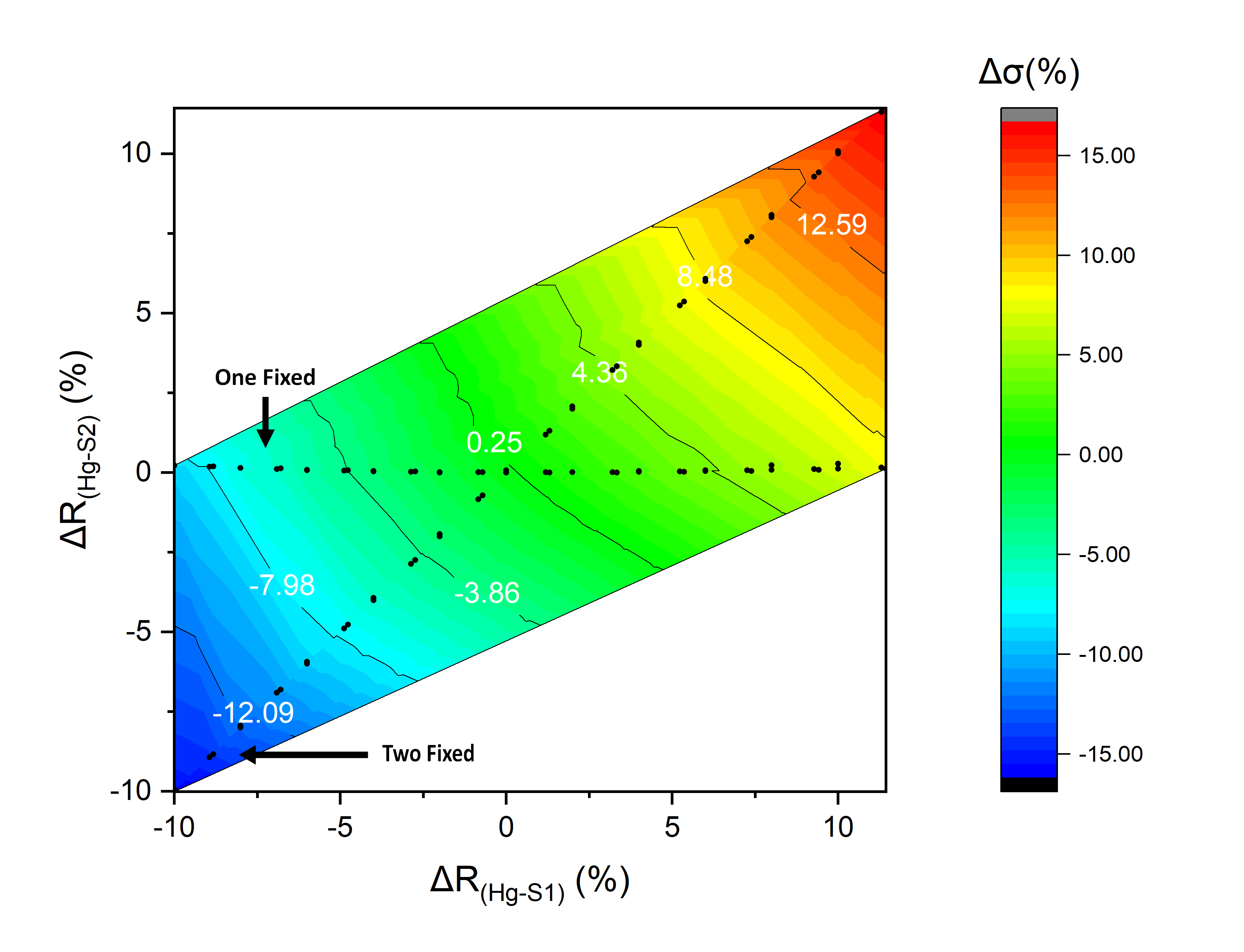}
    \caption{Contour plot of the effect of changes in the Hg-S bond lengths on the $\sigma$($^{199}$Hg) isotropic shielding constant in Hg(SR)$_2$ molecules calculated at ZORA/PBE0 level of theory.}
    \label{fig:3D_HgSR2}
\end{figure}

By plotting the previous results with $R_{\mathrm {(Hg-S1)}}$ as $x$ values and $R_{\mathrm {(Hg-S2)}}$ as $y$ values against $\sigma$($^{199}$Hg) as $z$, we can obtain a 3D model of how the $^{199}$Hg shielding constant is affected by the bond length in Hg(SR)$_{2}$ molecules. Fig. \ref{fig:3D_HgSR2} is thus the 3D version of the combination of Figs. \ref{fig:HgSH2_BL_zora},\ref{fig:HgSR2_1bl} and \ref{fig:HgSR2_2bl}. 

\begin{figure}[h]
    \centering
    \includegraphics[width=0.45\textwidth]{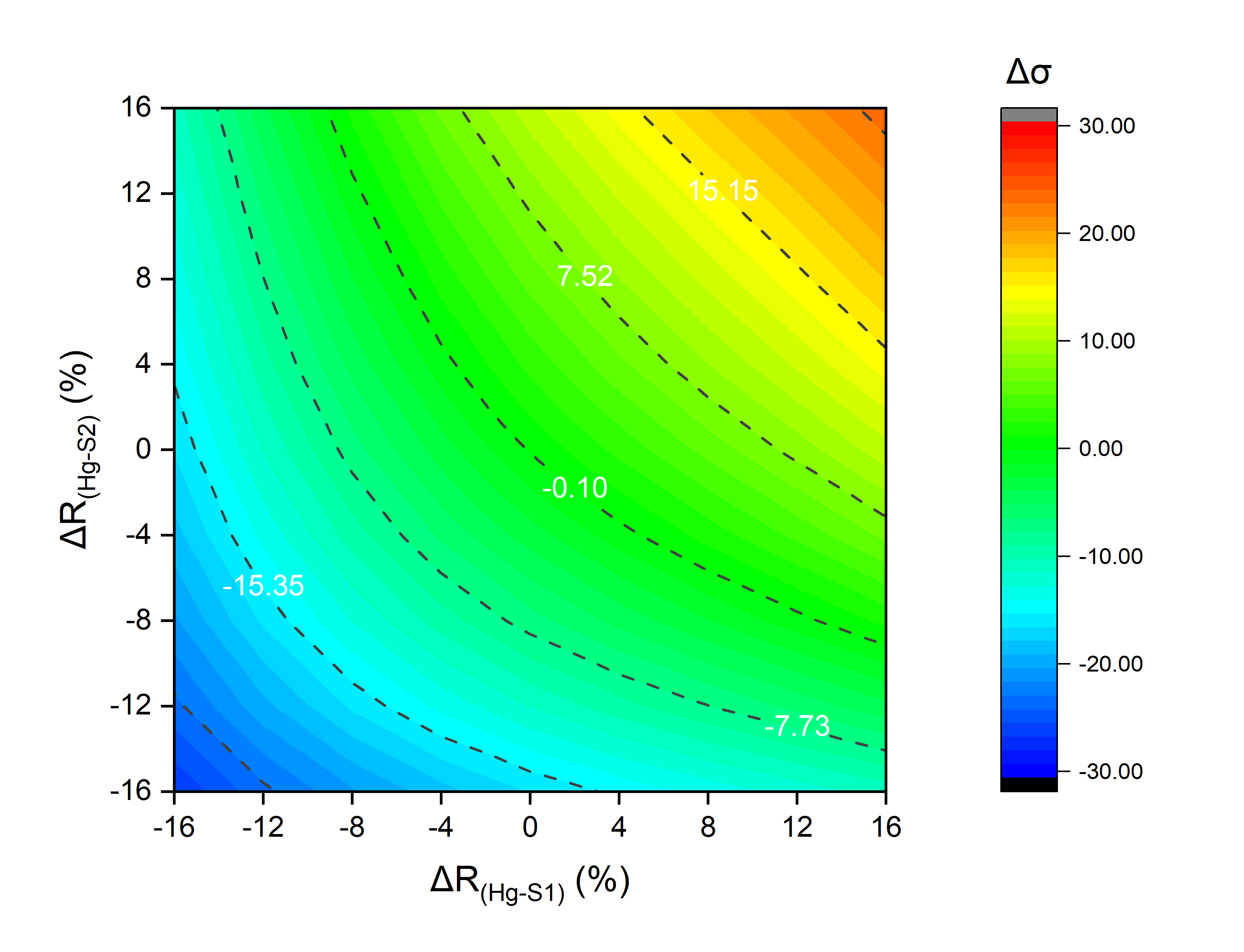}
    \caption{Hg(SMe)$_2$: Contour plot of the effect of changes in both Hg-S bond lengths on the $\sigma$($^{199}$Hg) isotropic shielding constant  calculated at ZORA/PBE0 level of theory.}
    \label{fig:grid_HgSMe2}
\end{figure}

To obtain a more complete picture, we took Hg(SMe)$_2$ as an example and made a whole grid of changes in both bond lengths.
The Hg-S1 and Hg-S2 bond lengths were modified independently from -20\% to 20\% by step of 4\%.
Furthermore, in the central area, $-2\sim2$\%, a denser grid was made with steps of 0.4\%.
The results are shown in Fig. \ref{fig:grid_HgSMe2}.

By fitting the surface in Fig. \ref{fig:3D_HgSR2} and Fig. \ref{fig:grid_HgSMe2} with a two dimensional polynomial model, the following equations can be derived for the change in the shielding constant, $\Delta\sigma$, for Hg(SR)$_2$,
\begin{equation}
\begin{aligned}
\Delta\sigma =& 0.7714x+0.7375y\\
    &-0.0107x^2-0.0156y^2+0.0200xy\\
R=&0.9994
\end{aligned}
\end{equation}
and for \SMed,
\begin{equation}
\begin{aligned}
    \Delta\sigma =&-0.0169+ 0.8203x+0.8178y\\
    &-0.0113x^2-0.0112y^2+0.0190xy\\
    R=&0.9992
\end{aligned}
\end{equation}
where $x$ and $y$ are the change of two bond lengths $ \Delta R_{\mathrm{(Hg-S1)}}$ and  $\Delta R_{\mathrm{(Hg-S2)}}$ and $R$ is the coefficient of determination.
For the grid of {\SMed}, the fitting model was constrained with $\Delta\sigma(0,0)=0$. 
For the Hg(SR)$_2$ fitting, a constraint of cross term coefficient $f=0.02$ was set. 
These constraints do not strongly affect the results, in contrast, they lower the dependency of parameter significantly. 
Comparing the two 2D fit functions, one observes that Hg(SR)$_2$ and {\SMed} give similar functions in terms of the two independent variables $\Delta R_{\mathrm{(Hg-S1)}}$ and  $\Delta R_{\mathrm{(Hg-S2)}}$.
A reason for the differences might be that for {\SMed} the bond lenghts were changed by up to -20\%, where the bond length becomes $\sim1.8$ \AA, which is very unlikely to appear in experiment. 
At these extreme geometries {\SMed} has strange $\sigma$($^{199}$Hg) values, which could be consider as data contamination. 

Thus, the leading terms in the range of physically realistic values of the bond lengths (x and y) are the first order terms. this implies that the effect on the isotropic shielding of changing the bond length of one bond or the other are largely independent of each other, and therefore additive.

\subsubsection{\SHt}
The coordination chemistry of mercury(II) is more complicated than just two-fold coordinated complexes. Therefore, we need to investigate, whether the conclusion from the previous section 
that $\sigma$($^{199}$Hg) increases with increasing Hg-S bond lengths, also holds for higher-coordinated complexes. 
This subsection will, therefore, focus on \SHt. Based on the other conclusion drawn in the previous section, i.e. that the side groups only slightly affect the geometry dependence of the shielding constant of the central mercury atom, \SHt ought to be representative enough for such complexes while also being computationally affordable. 
In this subsection, the same workflow as previous, i.e. to carry out shielding constant calculations on a series of constrained optimized geometries, will be followed. 
Constrained geometry optimization and shielding constant calculations were performed at the DFT/PBE0 level with the ZORA method. 
The QZ4P basis sets were employed for mercury and sulfur atoms, while TZ2P were for used carbon atoms and DZP on other atoms. 
4-component calculations were carried out also at the DFT/PBE0 theory level with the v3z basis sets on mercury and sulfur atoms and the v2z basis set on other atoms (see Figures S1 to S8 and Tables S1 to S8 in the supplementary material).

\begin{figure}[h!]
    \centering
    \includegraphics[width=0.45\textwidth]{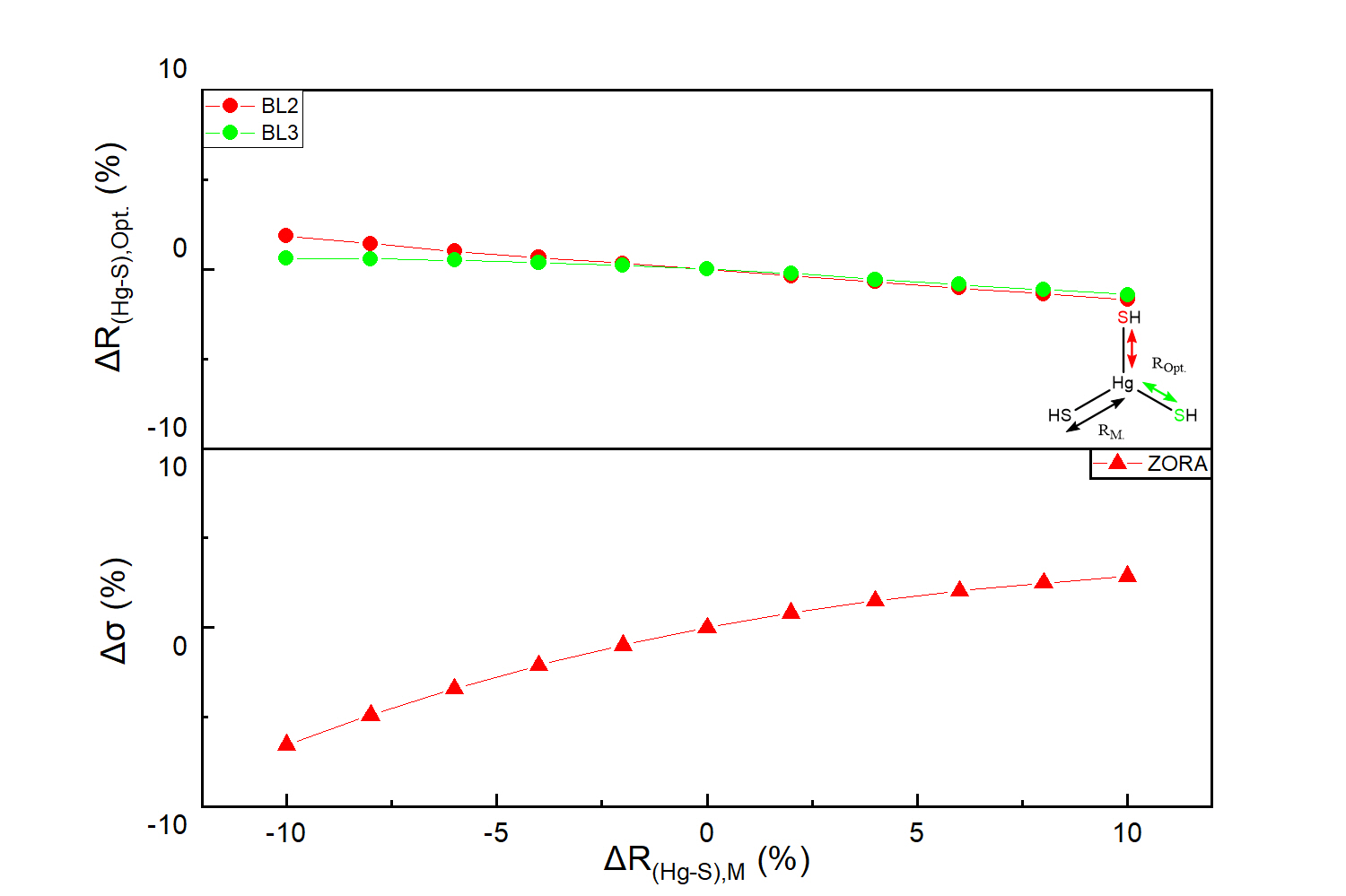}
    \caption{{\SHt}: $\sigma$($^{199}$Hg) isotropic shielding constant variation vs. changes in one Hg-S bond length calculated at the ZORA/PBE0/QZ4P level of theory.
    }
    \label{fig:HgSH3_1bl}
\end{figure}

\begin{figure}[h!]
    \centering
    \includegraphics[width=0.45\textwidth]{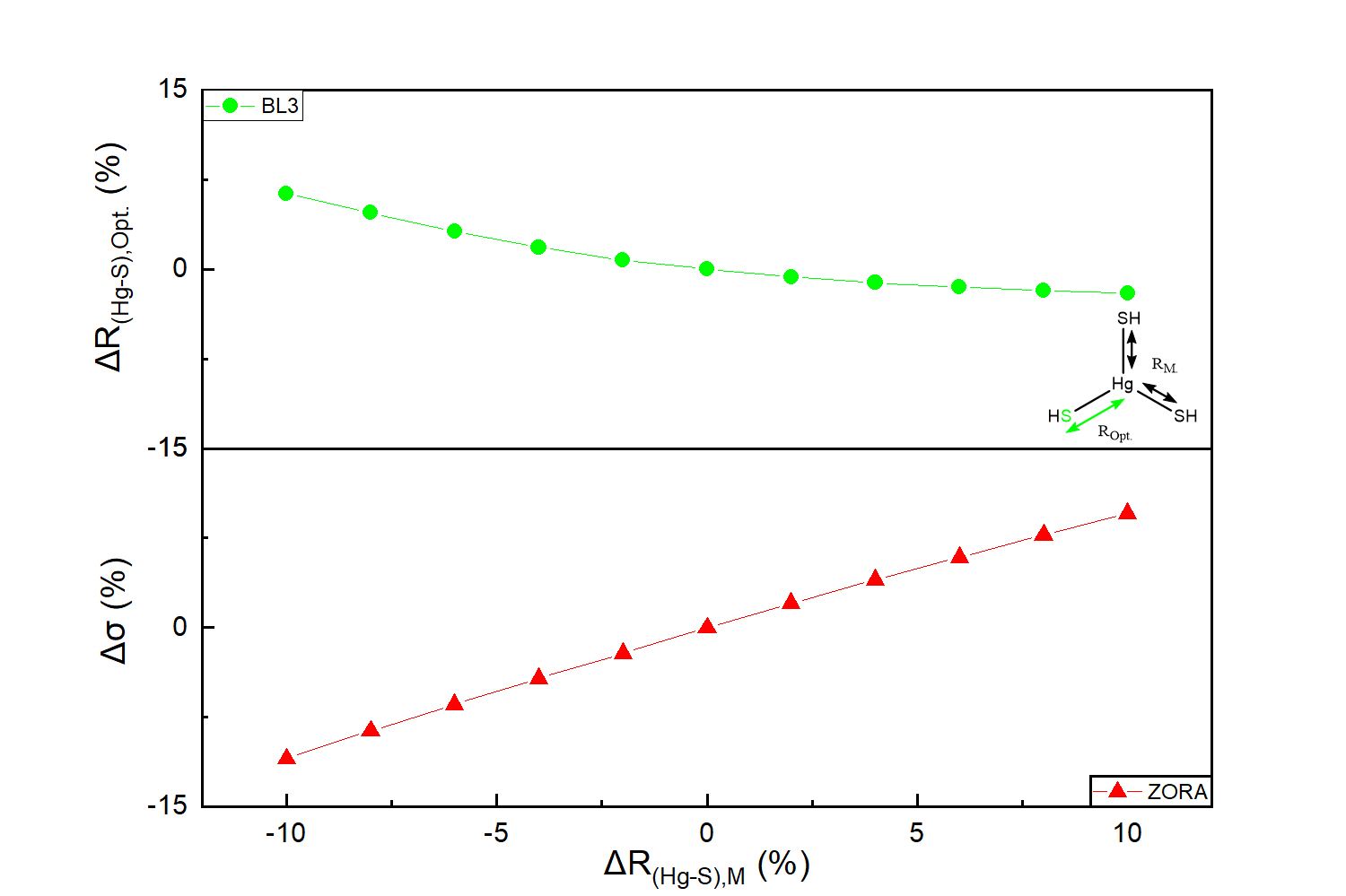}
    \caption{{\SHt}: $\sigma$($^{199}$Hg) isotropic shielding constant variation vs. changes in two Hg-S bond lengths in {\SHt} calculated at the ZORA/PBE0/QZ4P level of theory.}
    \label{fig:HgSH3_2bl}
\end{figure}

\begin{figure}[h!]
    \centering
    \includegraphics[width=0.45\textwidth]{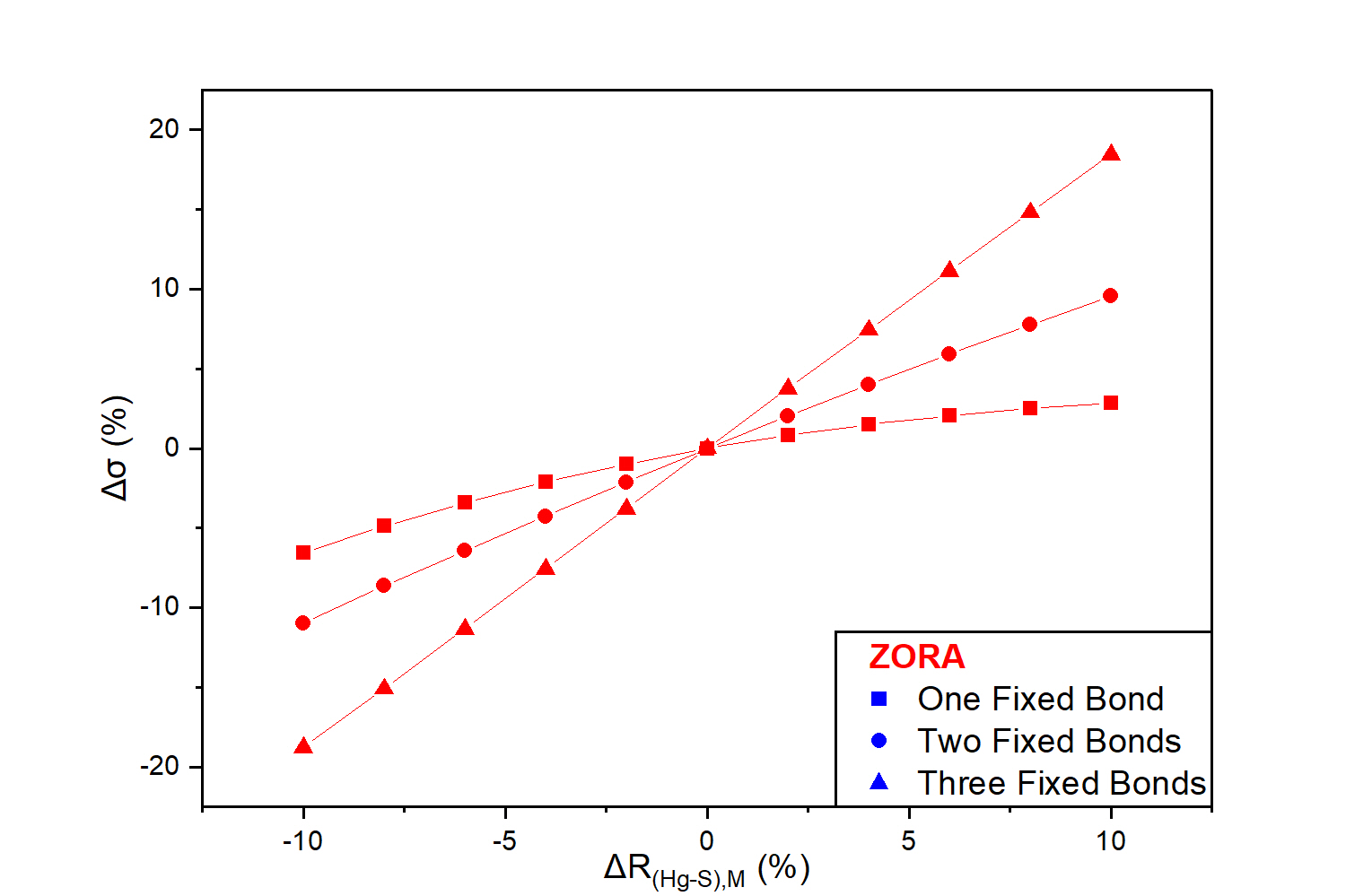}
    \caption{{\SHt}: $\sigma$($^{199}$Hg) isotropic shielding constant variation vs. changes in one, two or three Hg-S bond length calculated at the ZORA/PBE0/QZ4P level of theory.}
    \label{fig:HgSH3_3bl_4comp}
\end{figure}

In Fig. \ref{fig:HgSH3_1bl}, with the constraint of one bond fixed, the other two free bonds tend to compensate the modification after re-optimizing the geometry, although the changes are small. 
The two free bonds change conversely by $1.7\%\sim-1.7\%$ while the the fixed bond is again modified by -10\% to 10\%. 
When we constrain two bonds simultaneously, Fig. \ref{fig:HgSH3_2bl}, there is only one free bond, which again compensates the changes in the other bonds. 
However, here the length of the free bond changes by up to 6.3\% and thus more than when only one bond was fixed in Fig. \ref{fig:HgSH3_1bl}.
Looking now at the changes in $\sigma$($^{199}$Hg), when two (Fig. \ref{fig:HgSH3_2bl}) or all three (Fig. \ref{fig:HgSH3_3bl_4comp}) of the Hg-S bonds are varied, we observe again as for {\SHd} in Fig. \ref{fig:HgSH2_BL_zora} that changing more than one bond leads to larger changes in $\sigma$($^{199}$Hg). For all three bonds simultaneously varied by $\pm$10\% the changes in $\sigma$($^{199}$Hg) are up to $\pm$20\%.

\begin{figure}[h]
    \centering
    \includegraphics[width=0.45\textwidth]{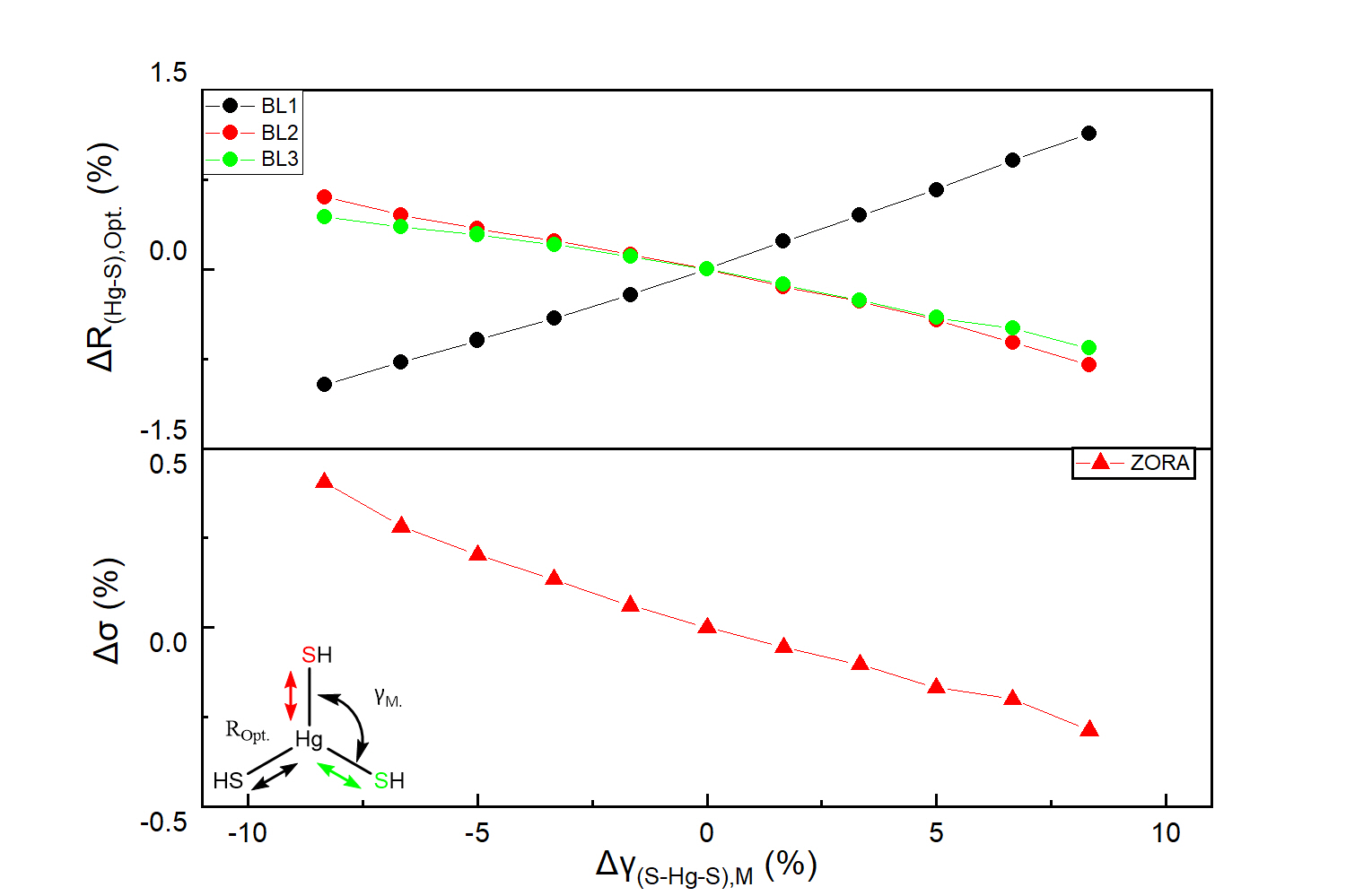}
    \caption{{\SHt}: $\sigma$($^{199}$Hg) isotropic shielding constant variation vs. changes in the S-Hg-S bond angle calculated at the ZORA/PBE0/QZ4P level of theory.}
    \label{fig:HgSH3_ba}
\end{figure}

In Fig. \ref{fig:HgSH3_ba}, the S-Hg-S angle $\Delta \mathrm{\gamma_{M}}$ was varied from 110$^\circ$ to 130$^\circ$. 
The bond lengths and the $^{199}$Hg shielding constant were not affected significantly but more than it was the case in {\SHd} in Fig. \ref{fig:HgSH2_ba}.
Again the re-optimized geometrical parameters, i.e. the three Hg-S bond lenghts compensate in a way the modification in the bond angle. 
The two adjacent bonds are shortened, when the angle increases, while the opposite bond length increases. 
Also $\sigma$($^{199}$Hg) changes more than in {\SHd} in Fig. \ref{fig:HgSH2_ba}, but it is not clear, whether this is a consequence of the change in the bond angle or in the consequently changes in the bond lengths.

\begin{figure}[h]
    \centering
    \includegraphics[width=0.45\textwidth]{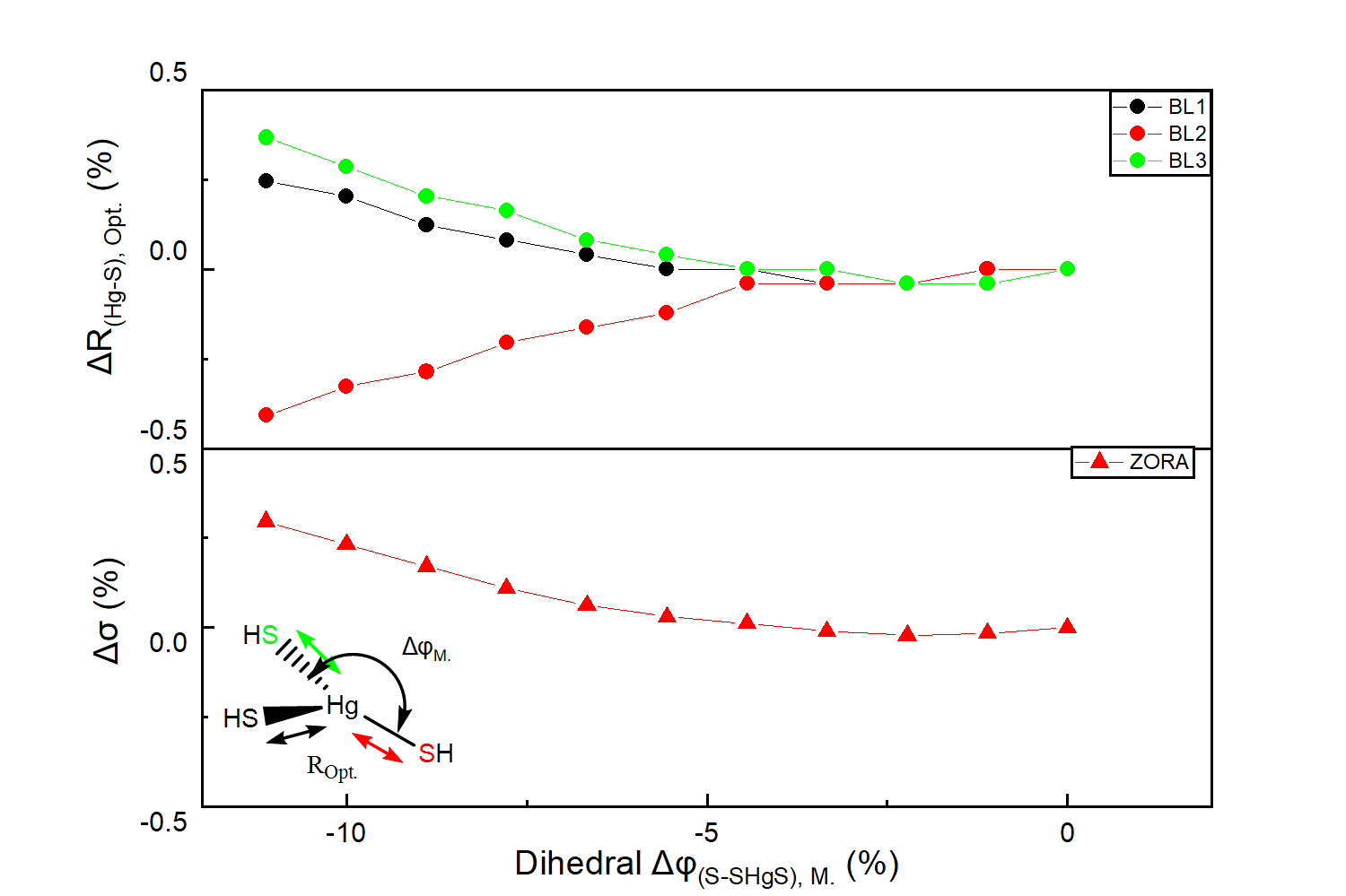}
    \caption{{\SHt}: $\sigma$($^{199}$Hg) isotropic shielding constant variation vs. changes in the S-SHgS dihedral angle calculated at the ZORA/PBE0/QZ4P level of theory.}
    \label{fig:HgSH3_di}
\end{figure}
Fig. \ref{fig:HgSH3_di}, finally, shows the effect of varying the S-SHgS dihedral angle and thus going away from a planar coordination around the central Hg atom.
The effect is even smaller than by changing the S-Hg-S bond angle. 
When we vary the dihedral from 180$^\circ$ to 160$^\circ$, the three bond lengths change slightly by $\pm0.5\%$. 
The maximum change of the shielding constant by varying the dihedral angle is below 0.3\%. 


\section{Conclusion}

We have studied how the $^{199}$Hg isotropic shielding constant in [Hg(SR)$_n$]$^{2-n}$ complexes changes, when we systematically vary bond lengths or bond angles and re-optimize the remaining internal coordinates.
Furthermore, we have investigated how the details of the computational method, i.e. the one-electron basis set, the DFT exchange-correlation functional or the method for treating relativistic effects, affect both the geometry optimizations and the calculations of the $^{199}$Hg isotropic shielding constant.

First of all, we can conclude that the geometries optimized with the ZORA method are very close to the ones obtained with 4-component relativistic calculations as long as large enough basis sets are employed, e.g. the results of an ZORA/PBE0/QZ4P calculations versus a 4-component/PBE0/v3z calculation, where the previous requires significantly less computational resources than the latter.

In the calculation of the $^{199}$Hg isotropic shielding constant for different complexes and many different geometries, ZORA with the QZ4P basis set consistently underestimates the results of 4-component/v3z calculations by ca. 1073 ppm. 
This implies than that both methods predict almost the same chemical shifts (within 55 ppm using large  basis sets) of Hg in [Hg(SH)$_n$]$^{2-n}$ compounds. 
For calculations of chemical shifts in larger systems, ZORA is thus an acceptable and much more cost-effective choice. 
The choice of XC-functionals, on the other hand, affects significantly the calculated geometries and chemical shifts. 

We find that varying the Hg-S bond lengths has a large effect on the $^{199}$Hg isotropic shielding constant in contrast to varying the S-Hg-S bond angles or even a dihedral angle.
On changing one of the Hg-S bond lengths by up to 10\% while re-optimizing the other bond(s) the percentage change in the shielding constant is almost as large.
Varying simultaneously two or three Hg-S bonds leads to even larger changes in the calculated shielding constants.
Increasing the bond lengths increases also the absolute shielding constants.
The side groups of coordinating thiolate in mercury complexes may affect the mercury absolute value of NMR shielding. 
However, they have virtually no influence on how much the $^{199}$Hg isotropic shielding constant changes when varying the bond lengths. The curves for the different compounds relating the percentage change in shielding constant to the percentage change in the modified Hg-S bond length are virtually identical.
For the series of [Hg(SR)$_n$]$^{2-n}$ complexes, the shielding constant of mercury is thus consistently nearly proportional to the Hg-S bond length. 
The change of S-Hg-S bond angle and S-SHgS dihedral does not influence the shielding constants much, and only influence the shielding constants by giving rise to bond length changes. 



\begin{acknowledgments}

\end{acknowledgments}

\section*{Data Availability Statement}
The data that support the findings of this study are available from the corresponding author upon reasonable request.

\providecommand{\noopsort}[1]{}\providecommand{\singleletter}[1]{#1}%

\end{document}